\documentclass[amsmath,amssymb,aps]{revtex4}

\selectfont 
\usepackage{hyperref}
\usepackage{amsmath}
\usepackage{graphicx}
\usepackage{bm}
\usepackage{braket}
\usepackage{tensor}
\usepackage{color}
\usepackage{slashed}
\usepackage{color}
\usepackage{booktabs} 
\usepackage{multirow}
\usepackage{float}
\usepackage{kantlipsum} 

\def\omegat{\widetilde{\omega}}

\newcommand{\beq}{\begin{equation}}
\newcommand{\eeq}{\end{equation}}
\newcommand{\beqy}{\begin{eqnarray}}
\newcommand{\eeqy}{\end{eqnarray}}
\def\be{\begin{equation}}
\def\ee{\end{equation}}
\def\bea{\begin{eqnarray}}
\def\eea{\end{eqnarray}}

\begin{document}

\newcommand{\rp}[1]{\textcolor{blue}{AU: #1}}

\title{Some Observable Physical Properties of the Higher Dimensional dS/AdS Black Holes in Einstein-Bumblebee Gravity Theory}

\author{Akhil Uniyal}
\email{akhil\_uniyal@iitg.ac.in}

\affiliation{Department of Physics, Indian Institute of Technology, Guwahati 781039, India}

\author{Sara Kanzi}
\email{sara.kanzi@final.edu.tr}

\affiliation{Faculty of Engineering, Final International University,
Kyrenia 99370, North Cyprus via Mersin 10 Turkey}

\author{\.{I}zzet Sakall{\i}}
\email{izzet.sakalli@emu.edu.tr}

\affiliation{Physics Department, Eastern Mediterranean
University, Famagusta 99628, North Cyprus via Mersin 10, Turkey}

\date{\today}

\begin{abstract}
We study the greybody factors, quasinormal modes, and shadow of the higher dimensional de-Sitter (dS)/ anti de-Sitter (AdS) black hole spacetimes derived from the Einstein-bumblebee gravity theory within the Lorentz symmetry breaking (LSB) framework. We specifically apply the semi-analytical WKB method and the time domain approach to study the scalar and Dirac perturbations of the black hole. In-depth researches are done on the effects of the LSB and dimensionality on the bosonic/fermionic greybody factors, quasinormal modes, and shadow of the higher dimensional bumblebee black hole. The results obtained are discussed, tabulated, and illustrated graphically.

\end{abstract}

\pacs{nnnnn}

\maketitle

%
%
%

\section{Introduction}\label{sec:intro}
Since the Standard Model (SM) of particle physics \cite{ParticleDataGroup:2012pjm} and older theories like general relativity (GR) \cite{Weinberg:1988cp}, which describes how matter warps spacetime, cannot explain everything in the universe, including what occurs in the vicinity of a black hole, physicists are continually working to develop new and better ideas. Investigating that any retained concept such as Lorentz symmetry \cite{Hehl:1994ue} may not be true in extreme cases is a very fruitful approach to explore for new physics. According to some gravitational wave models, the cosmos is not entirely symmetrical. Because of these ideas, the cosmos will always have extra elements that prevent it from perfectly adhering to the Lorentz symmetry. In other words, the cosmos would have a unique or favored orientation. These new models explain a theory known as "bumblebee gravity (BG)" \cite{Bailey:2006fd,Kostelecky:2010ze,Bluhm:2004ep}. Its name comes from the alleged remark made by experts that bumblebees should not be allowed to fly since we did not know how their wings produced lift. Specifically, we do not fully comprehend how these gravity theories function and how they may be consistent with the universe that we currently observe. The possible use of bumblebee gravity models to explain dark energy \cite{ParticleDataGroup:2018ovx}, the phenomena that causes the cosmos to expand at an accelerated rate \cite{Copeland:2006wr}, is one of their most effective applications. It turns out that an effect that causes our universe to expand faster can be related to how much Lorentz symmetry our universe breaches. In addition, this notion seems very enticing because we do not know what is generating dark energy. In short, this is the direction that the bumblebee gravity theory is anticipated to contribute to GR and subsequently to the quantum gravity theory (QGT) \cite{DeWitt:1967yk,Aharony:1999ti}.

Lorentz violations (LVs) affect both the predictions of the SM of particle physics and the basic predictions of special relativity, including the concept of relativity, the constancy of the speed of light in all inertial frames of reference, and time dilation. Test theories for special relativity and effective field theories, like the Standard Model Extension (SME) \cite{Jungman:1995df}, have been developed to evaluate and forecast any violations \cite{Reyes:2021cpx}. SME relates the SM to GR and includes additional features such as the LVs operating at the Planck scale \cite{Liberati:2009pf}. In other words, SME is an effective field theory integrating GR and SM on low energy scales. Other theories that propose the LVs besides SME include string theory \cite{Kostelecky:1988zi}, Einstein-aether theory \cite{Jacobson:2000xp,Jacobson:2004ts}, non-commutative field theory \cite{Carroll:2001ws,Mocioiu:2000ip,Ferrari:2006gs}, loop QGT \cite{Gambini:1998it,Ellis:1999uh}, brane-world scenarios \cite{Burgess:2002tb,Frey:2003jq}, and massive gravity \cite{Fernando:2014gda}, also check \cite{Reyes:2022mvm}. For the purpose of examining potential visible signs of a breach of particle Lorentz symmetry, the SME is experimentally accessible. In the model of SME, a spontaneous symmetry breaking potential caused by self-interacting tensor fields having vacuum expectation values (VEV), yields to the background tensor fields, which provides the local LV. An example of such a particular theory is the bumblebee field $B_{\mu}$, a self-interacting tensor field with a non-zero VEV with $b_{\mu}$ that specifies a preferred direction in spacetime and spontaneously breaks the Lorentz symmetry. The potential of a bumblebee field can take many different forms. Among the other bumblebee's potentials, $V(X)=k X^{2} / 2$ is a smooth and functional potential, where $k$ is a constant \cite{Bluhm:2007bd} and $X=B^{\mu} B_{\mu} \pm b^{2}$. When the bumblebee field matches its VEV, it has a minimum: $V=0$ and $V^{\prime}(X)=$ $k X=0$  when $X=0$. With this particular potential, the static black hole solutions in the Einstein-bumblebee gravity theory (EBGT)  were derived by Bertolami et al. \cite{Bertolami:2005bh} and by Casana et al. \cite{Casana:2017jkc}. It was revealed by \cite{Ding:2020kfr,Maluf:2022knd,Kanzi:2022vhp} that the solution obtained by  Ding et al. \cite{Ding:2019mal}, who claimed to have found a rotating black hole in the BG gravity model, was actually wrong. The first physically accepted \textit{slowly} rotating black hole solution in the EBGT was wrong. The first physically accepted \textit{slowly} rotating black
hole solution in the EBGT was obtained by Ding and Chen \cite{Ding:2020kfr}. Subsequently, Jaha et al. \cite{Jha:2020pvk} and  Poulis and Soares \cite{Poulis:2021nqh} have recently managed to derive an arbitrarily spinning
bumblebee black hole solution, which means crossing a difficult threshold in the EBGT. Another milestone
in this regard was passed by \cite{Ding:2022qcy}, who achieved an exact higher dimensional anti-de Sitter (AdS) black hole
solution in the EBGT. This AdS black hole can only exist with a special bumblebee potential having a
linear functional form with a Lagrange-multiplier field $\lambda$. It is worth noting that, this additional field can be
absorbed by the construction of an effective cosmological constant $\Lambda_{e}$ and is rigidly restricted by the equation
of bumblebee motion. Furthermore, the obtained higher dimensional black hole of the EBGT is nothing but a
Schwarzschild-AdS-like black hole solution since it cannot asymptotically approach anti-de Sitter spacetime,
just as the Schwarzschild-like black hole \cite{Casana:2017jkc}.
The bumblebee field has an impact on the black hole horizon location, in contrast to the Schwarzschild-like
black hole \cite{Casana:2017jkc}. For the higher dimensional bumblebee metric, we compute very important observables in order to relate
the bumblebee field to the spacetime geometry: the greybody factors (GbFs), quasinormal modes (QNMs),
and shadow angular radius. Let us now briefly recall what the observables in question are: GbFs, which
distinguish a black hole’s thermal emission spectrum from a pure black-body spectrum, are functions of
frequency, angular momentum, and black hole parameters. In other words, GbF is a quantity related to the
quantum nature of a black hole, and there are different approches to computing the GbF \cite{sarak1,sarak2,sarak3,sarak4,sarak5,sarak6,sarak7,sarak8}. \\
The modes of energy dissipation of a perturbed black hole or field are known
as QNMs, and they characterize the perturbations of a field that dissipates with time. The solutions of the
relevant perturbation equations that fulfill the boundary conditions necessary for purely incoming waves at
the horizon and purely outgoing waves at infinity are represented by a black hole’s QNMs. One must obtain
the gravitational QNMs’ spectra in order to examine the stability of higher-dimensional black hole solutions
that might be present in nature. The stable and unstable black holes have a relationship with the damped
and undamped states, respectively. Numerous techniques have been used to examine the QNMs frequencies, including the analytical method \cite{Chandrasekhar:1984siy,Ovgun:2017dvs,Jusufi:2017trn,Sakalli:2021dxd,Sakalli:2016fif,Sakalli:2018nug}, WKB method \cite{sarak2,Schutz:1985km,Iyer:1986np,Iyer:1986nq}, Frobenius method \cite{Konoplya:2011qq}, continuous fractions method
\cite{Leaver:1985ax}, Mashhoon method \cite{izmashoon}, feedforward neural network method \cite{Ovgun:2019yor}, and many more (for topical reviews, the reader is referred to \cite{Kokkotas:1999bd,Berti:2009kk,Sakalli:2022xrb}). In our work,
we mainly focus on the WKB approximation method \cite{Schutz:1985km,Iyer:1986np,Iyer:1986nq} to compute the QNMs. The QNMs
up to third order were first computed by Iyer and Will \cite{Iyer:1986np}. Then, Konoplya \cite{Konoplya:2003ii} made it possible for
us to compute the QNMs frequencies without using laborious numerical techniques, this resulted in a higher
order contribution.\\ 
Recently, the shadows of black holes have become one of the primary issues in physics.
The reason for that is the Event Horizon Telescope Collaboration’s debut photograph of a black hole, which
was first released in 2019 \cite{EHT:2019nmr,EventHorizonTelescope:2019dse,EventHorizonTelescope:2019ths}. In fact, those pictures captured with the cutting-edge technology depict the shadow of M87 \cite{EventHorizonTelescope:2019pgp} and SgrA* \cite{EventHorizonTelescope:2022xnr,Dokuchaev:2020wqk},
which are the supermassive black holes in the galaxy M87 and in the Milky Way Galaxy, respectively. But it was only during the last century that the first black hole’s shadow was estimated. Synge \cite{Synge:1966okc}
acquired what is known as the shadow of the Schwarzschild black hole today in the 1960s. Bardeen \cite{Bardeen:1973tla}, thereafter
extended Synge's work to the Kerr geometry. Recently, shadows have been considered for a number of
black holes in a variety of scenarios. Recent researches suggest that there might be relationships between the
shadow and black hole properties in general relativity or even in contexts outside of the Einsteinian paradigm \cite{Neves:2019lio,Vagnozzi:2019apd,Kumar:2020yem,Uniyal:2022vdu}. Shadow for the slowly rotating Kerr-like black hole in Einstein bumblebee gravity has been studied \cite{Kuang:2022xjp}. We will be looking the effect of bumblee gravity on the shadow in higher dimension dS/AdS spacetime.

The main purpose of this paper is to study the perturbations of scalar and fermion fields in the higher dimensional dS/AdS black hole geometries of the EBGT and the shadows of those higher dimensional black holes. For the perturbations, we shall consider the Klein-Gordon and Dirac equations. The obtained wave equations allow for semi-analytical methods to be used for GbF and QNM analyses. Then, we consider the photon’s orbit and radius of the shadow of the black hole. This article is structured as follows: In Sec. \ref{sec:formal}, we briefly introduce the higher dimensional dS/AdS black hole solutions in the EBGT \cite{Ding:2022qcy}. Then, we discuss the GbFs of bosons in Sec. \ref{sec:gamma}. Section \ref{sec:waveEq} is devoted to Dirac equation of massless fermions on the higher dimensional dS/AdS black hole spacetime. We also compute the rigorous lower bounds on the fermionic GbFs. Then, the QNMs are studied in Sec. \ref{sec:QNMs}. Null geodesics and shadow radius problems are discussed in Sec. \ref{sec:Shadow Radius}. The purpose of Sec. \ref{sec:Relations} is to investigate the connection between shadow radius and QNMs. In Sec. \ref{sec:cnstraint}, we check the effect of the bumblebee parameter on the shadow diameter via the real black holes. We draw our conclusions in Sec. \ref{sec:conclude}.

\section{Higher dimensional \lowercase{d}S/A\lowercase{d}S black holes in EBGT} \label{sec:formal}

The bumblebee vector field $B_{\mu}$ in the EBGT, has included a vacuum expectation value which is nonzero, in order to define a unconstrained Lorentz symmetry breaking via a given potential. The action of Einstein-bumblebee gravity in higher dimensions $D \geq 4$ is given by \cite{Ding:2022qcy, Ding:2020kfr},

\begin{equation}
S=\int d^Dx \sqrt{-g} [\frac{R-2\Lambda}{2\kappa}+\frac{\varrho}{2\kappa}B^{\mu}B^{\nu}R_{\mu \nu}-V(B_{\mu}B^{\nu} \mp b^2)+\mathcal{L}_M],\label{1}
\end{equation}

where, $\Lambda$ is the cosmological constant. $\kappa=8\pi G_D/c^4$, where $G_D=G\Omega_{D-2}/4\pi$ \cite{Boulware:1985wk} and $\Omega_{D-2}=2\sqrt{\pi}^{D-1}/\Gamma[(D-1)/2]$ is the area of a unit $D-2$ sphere. From now on we will take $G_D=1$ and $c=1$ for simplicity. $b$ is a positive constant and $\mathcal{L}_M$ represents the matter Lagrangian form.\\
The strength of the non-minimal coupling of gravity with the bumblebee field $B_{\mu}$ is determined by the coupling constant $\varrho$. The potential term $V(B_{\mu}B^{\nu} \mp b^2)$ represents Lorentz or CPT (charge, parity and time) violation. The potential has a minima at $B^\mu B_\nu \pm b^2 =0$ and $V'(b_\mu b^\mu)=0$, which destroy the $U(1)$ symmetry. The bumblebee field $B_\mu$ takes a nonzero vacuum expectation value (VEV) $<B^\mu>=b^\mu$ at these minima which tells us that vacuum of this model has a preferred direction in the spacetime. The vector $b^\mu$ here is a constant function of spacetime which has a value $b_\mu b^\mu= \mp b^2$, where the $\pm$ signs denote timelike or spacelike forms of the vector $b^\mu$. The bumblebee field strength is given by

\begin{equation}\label{2}
B_{\mu \nu}=\partial_\mu B_\nu - \partial_\nu B_\mu.
\end{equation}

We get the following constraint on the $B_{\mu \nu}$ due to its antisymmetric nature \cite{Bluhm:2007bd},

\begin{equation}\label{3}
\nabla^\mu \nabla^\nu B_{\mu \nu}=0.
\end{equation}

Varying the action \eqref{1} with respect to the metric yields the following field equation:

\begin{equation}\label{4}
G_{\mu \nu}+\Lambda g_{\mu \nu}=\kappa T^B_{\mu \nu}+\kappa T^M_{\mu \nu},
\end{equation}

where $G_{\mu \nu}=R_{\mu \nu}-g_{\mu \nu}R/2$ and $T^B_{\mu \nu}$ is known as the bumblebee energy momentum tensor, which is expressed by

\begin{multline}\label{5}
T^B_{\mu \nu}=B_{\mu \alpha} B^\alpha_\nu-\frac{1}{4} g_{\mu \nu} B^{\alpha \beta}B_{\alpha \beta}-g_{\mu \nu} V+2B_\mu B_\nu V' \\ +\frac{\varrho}{\kappa} \left[ \frac{1}{2} g_{\mu \nu} B^\alpha B^\beta R_{\alpha \beta}-B_\mu B^\alpha R_{\alpha \nu} \right]\\ +\frac{\varrho}{\kappa} \left[ \frac{1}{2} \nabla_\alpha \nabla_\mu (B^\alpha B_\nu)+\frac{1}{2} \nabla_\alpha \nabla_\nu (B^\alpha B_\mu)-\frac{1}{2} \nabla^2 (B^\mu B_\nu)-\frac{1}{2} g_{\mu \nu} \nabla_\alpha \nabla_\beta (B^\alpha B^\beta) \right].
\end{multline}

In the above expression, $V^{\prime}$ represents the differentiation of $V$ computed at $x=B^\mu B_\mu \pm b^2$. Then, varying the action \eqref{1} with respect to $t$, the bumblebee field gives the following field equation by assuming that there is no coupling between the bumblebee field and the Lagrangian of matter:

\begin{equation}\label{6}
\nabla^\mu B_{\mu \nu}=2V' B_\nu - \frac{\varrho}{\kappa} B^\mu R_{\mu \nu}.
\end{equation}

We now suppose that there is no matter field and the bumblebee field is frozen at its VEV. Namely, we have (see, for example,  \cite{Bertolami:2005bh, Casana:2017jkc})
\begin{equation}\label{7}
B_\mu=b_\mu.
\end{equation}
Now, since we intend to include the cosmological constant in our theory, the non-zero cosmological constant requires the linear form of the potential as being stated in \cite{Ding:2022qcy}:

\begin{equation}\label{8}
V=\frac{\lambda}{2}(B_\mu B^\mu-b^2),
\end{equation}

where, $\lambda$ is a non-zero constant and considered as a Lagrange-multiplier field. The potential vanishes for condition \eqref{7} and the derivative of the potential $V'=\frac{\lambda}{2}$ modifies the Einstein field equation. Hence, Eq. \eqref{4} recasts in \cite{Ding:2022qcy, Ding:2020kfr},

\begin{equation}\label{9}
G_{\mu \nu}=\kappa (\lambda b_\mu b_\nu +b_{\mu \alpha} b^{\alpha_\nu}-\frac{1}{4} g_{\mu \nu}b^{\alpha \beta}b_{\alpha \beta})+\varrho \left( \frac{1}{2} g_{\mu \nu}b^\alpha b^\beta R_{\alpha \beta}-b_\mu b^\alpha R_{\alpha \nu}-b_\mu b^\alpha R_{\alpha \nu}+\bar{B}_{\mu \nu} \right),
\end{equation}

where,

\begin{equation}\label{10}
\bar{B}_{\mu \nu}=\frac{\varrho}{2} \left[ \nabla_\alpha \nabla_\mu (b^\alpha b_\nu)+\nabla_\alpha \nabla_\nu (b^\alpha b_\mu)-\nabla^2(b_\mu b_\nu)-g_{\mu \nu} \nabla_\alpha \nabla_\beta (b^\alpha b^\beta) \right].
\end{equation}

Now, we would like to construct a $D$-dimensional static and spherically symmetric metric in the EBGT. To this end, let us consider the following metric anstaz:

\begin{equation}\label{11}
ds^2=-e^{2\phi(r)} dt^2+e^{2\psi (r)} dr^2+r^2 d\Omega^2_{D-2}.
\end{equation}

Since, the spacetime considered has a strong radial variation compared to the temporal changes, we consider that the bumblebee field has a radial finite vacuum expectation value. Hence, the spacelike bumblebee field turns out to be

\begin{equation}\label{12}
b_\mu=(0,b e^{\psi(r)},0,0....,0),
\end{equation}

where, $b$ is a positive constant. The bumblebee field strength is defined by

\begin{equation}\label{13}
b_{\mu \nu}=\partial_\mu b_\nu - \partial_\nu b_\mu,
\end{equation}

whose components and their divergences are now all zero. Therefore,

\begin{equation}\label{14}
\nabla^\mu b_{\mu \nu}=0.
\end{equation}

From Eq. \eqref{6}, we can see the projection of the Ricci tensor along the bumblebee field:

\begin{equation}\label{15}
b^\mu R_{\mu \nu}=\frac{\kappa \lambda}{\varrho}b_\nu.
\end{equation}

Using Eq. \eqref{9}, we will have three independent equations:

\begin{equation}\label{16}
(D-2)(1+L)[2r\psi'-(D-3)]+e^{2\psi}[(D-2)(D-3)-2\Lambda r^2]=0,
\end{equation}

\begin{multline}\label{17}
2Lr^2(\phi''+\phi'^2-\phi' \psi')-2L(D-2)r(\psi'+\phi')-2(D-2)r\phi' \\+e^{2\psi}[(D-2)(D-3)+2\kappa \lambda b^2 r^2-2\Lambda r^2]-(1+L)(D-2)(D-3)=0,
\end{multline}

\begin{multline}\label{18}
(1+L)[r^2(\phi''+\phi'^2-\phi' \psi')+\frac{(D-3)(D-4)}{2}+(D-3)r(\phi'-\psi')]\\+e^{-2\psi} \left [ \Lambda r^2-\frac{(D-3)(D-4)}{2} \right]=0,
\end{multline}

where the prime symbol denotes differentiation of a function with respect to its argument and the Lorentz-violating parameter is given by $L=\varrho b^2\geq 0$. Now, Eq. \eqref{16} leads to the following metric function:

\begin{equation}\label{19}
e^{2\psi}=\frac{1+L}{f(r)},
\end{equation}

where,

\begin{equation}\label{20}
f(r)=1-\frac{16\pi M}{(D-2)\Omega_{D-2}r^{D-3}}-\frac{2\Lambda}{(D-1)(D-2)}r^2,
\end{equation}

where $M$ is the mass of the black hole. To have the Schwarzchild-like solution \cite{Casana:2017jkc} for $\Lambda=0$, we set

\begin{equation}\label{21}
e^{2\phi}=f(r).
\end{equation}

Hence, the bumblebee field reads

\begin{equation}\label{22}
b_\mu=(0,b\sqrt{(1+L)/f(r)},0,0,....,0),
\end{equation}

and from Eqs. \eqref{15} and \eqref{18}, one can see that the following expression for the cosmological constant should hold:

\begin{equation}\label{23}
\Lambda=\frac{(D-2)\kappa \lambda}{2 \varrho}(1+L),
\end{equation}

which puts a constraint on the parameter $\lambda$ from the potential \eqref{8}. Therefore, $\lambda$ is not a new degree of freedom in the theory. Moreover, one can define an effective cosmological constant $\Lambda_e$ as follows

\begin{equation}\label{24}
\Lambda_e=\frac{(D-2)\kappa \lambda}{2\varrho},
\end{equation}

which means that $\Lambda=(1+L)\Lambda_e$. After all those computations, we get the final form of the metric as
\begin{equation}\label{25}
ds^2=-f(r) dt^2+\frac{1+L}{f(r)} dr^2+r^2 d\Omega^2_{D-2},
\end{equation}

where the metric function reads

\begin{equation}\label{26}
f(r)=1-\frac{16\pi M}{(D-2)\Omega_{D-2}r^{D-3}}-\frac{2(1+L)\Lambda_e}{(D-1)(D-2)}r^2.
\end{equation}

\begin{figure*}[hbt!]
  \includegraphics[width=.45\textwidth]{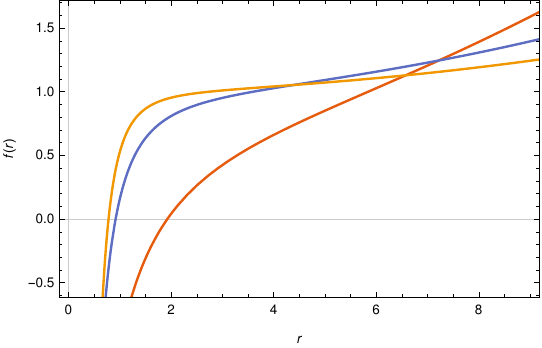}
  \includegraphics[width=.45\textwidth]{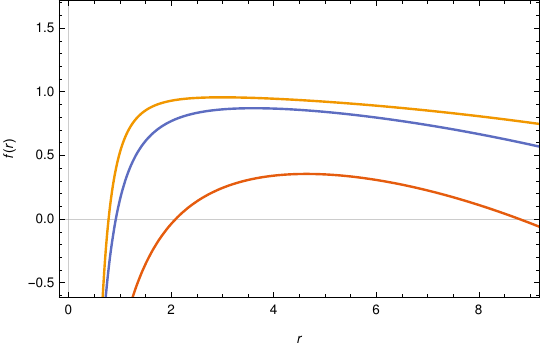}
\caption{Graph of the metric component $f(r)$ versus $r$ for various dimensions: $D=4$ (red), $D=5$ (blue), and $D=6$ (yellow). The physical parameters are chosen as $L=2, M=l=1$ and $\Lambda_e=-0.01$ (left), $\Lambda_e=0.01$ (right).} \label{fig:1a}
\end{figure*}

\begin{figure*}[hbt!]
  \includegraphics[width=.45\textwidth]{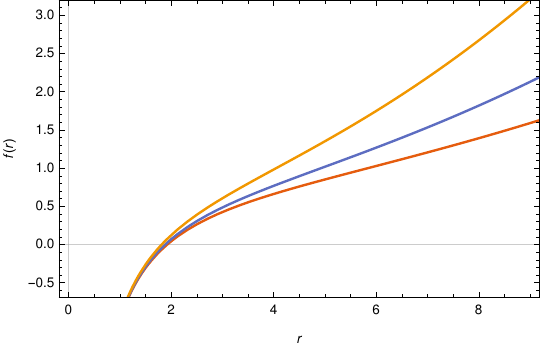}
  \includegraphics[width=.45\textwidth]{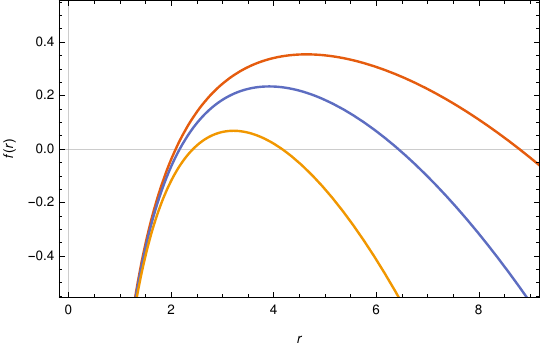}
\caption{Graph of the metric component $f(r)$ versus $r$ for various bumblebee parameters: $L=2$ (red), $L=4$ (blue), and $L=8$ (yellow). The physical parameters are chosen as $D=4, M=l=1$ and $\Lambda_e=-0.01$ (left), $\Lambda_e=0.01$ (right).} \label{fig:2a}
\end{figure*}

It is clear from the metric function \eqref{26} that the event horizon is affected by the bumblebee field. The behavior of the metric function is illustrated in Fig. \eqref{fig:1a} for both dS and AdS spacetimes with different dimensions. One can observe that we have one horizon (event) for the AdS case however for the dS case double horizons appear: event horizon (inner) and cosmological (outer) horizon (see Fig. \ref{fig:1a}). It is worth noting that we have found the similar kinds of behaviors for the varying LSB parameter, which are depicted in Fig. \ref{fig:2a}, in which $D=4$ is fixed for both for AdS and dS cases.

\section{G\lowercase{b}F\lowercase{s} \lowercase{of} BOSONS}\label{sec:gamma}

Among the various methods for deriving the GbFs of the relevant black hole spacetimes, only a small number of cases renders possible to derive their precise analytical expressions. In this section, we will apply the rigorous bounds technique to procure the GbFs of higher dimensional dS/AdS black hole in EBGT. To this end, the excitation of uncharged and massless scalar fields is going to be determined by the Klein-Gordon equation:
\begin{equation}
\Box \Psi=0, \label{s8}
\end{equation}
where $\Box$ denotes the D’Alembert operator. So, Eq. \eqref{s8} can be rewritten as
\begin{equation}
\begin{split}
\frac{1}{\sqrt{-g}}\partial_\mu(\sqrt{-g}g^{\mu \nu}\partial_\nu)\Psi=0, \label{s9}
\end{split}
\end{equation}

in which for our $D$-dimensional metric \eqref{25} $\sqrt{-g}$ is given by
\begin{equation}
\sqrt{-g}=r^{D-2} \sqrt{(1+L)}  \prod_{i=1}^{D-2} sin \theta_i. \label{s10}
\end{equation}
In order to get separate radial and angular Klein-Gordon equations, let us apply the following ansatz \cite{Harmark:2007jy}: 

\begin{equation}
\Psi=e^{-i \omega t} \phi(r) Y_{lm}(\Omega), \label{s11}
\end{equation}
where $\omega$ indicates frequency, $l$ represents the azimuthal quantum number, and $m (-l\leq m\leq l)$ is the spherical harmonic index. The angular equation yields the eigenvalue ($\lambda$) \cite{Sakalli:2011zz} as 
\begin{equation}
\begin{split}
\lambda=-l(D+l-3).
\end{split}
\end{equation}
Using all of the above equations we can write the radial equation for the scalar field,

\begin{equation}
\begin{split}
\phi^{''}+\left( \frac{D-2}{r}+ \frac{f^{'}}{f} \right) \phi^{'}++\left( \frac{\omega^2 (1+L)}{f^2}-\frac{(1+L)(l(D+l-3))}{f r^2} \right) \phi=0,
\end{split} \label{izss}
\end{equation}

where a prime mark denotes a derivative with respect to the radial coordinate, $r$. Applying the following transformation 
\begin{equation}
\phi=\frac{u}{r^{\frac{D-2}{2}}}, 
\end{equation} 

one can rewrite Eq. \eqref{izss} as

\begin{equation}
\begin{split}
\frac{f^2}{1+L}u^{''}+\frac{f f^{'}}{1+L}u^{'}+\left[\omega^2- \left( \frac{D-2}{2} \right) \frac{f f^{'}}{1+L} \frac{1}{r}-\frac{(D-2)(D-4)}{4r^2}\frac{f^2}{1+L}-\frac{l(D+l-3)f}{r^2} \right]u=0.
\end{split}
\end{equation}

At this stage, by using the tortoise coordinate $dr_{*}=\sqrt{1+L} \frac{dr}{f}$, we get a Schr\"{o}dinger-like wave equation:

\begin{equation}
\begin{split}
\frac{d^2u}{dr_*^2}+\left[\omega^2-V_{eff} \right]u=0, \label{sk11}
\end{split}
\end{equation}

where

\begin{equation}
\begin{split}
V_{eff}=f\left[\frac{(D-2)(D-4)}{4r^2}\frac{f}{1+L}+\frac{(D-2)f^{'}}{2(1+L)r}+\frac{l(D+l-3)}{r^2} \right]. \label{SS1}
\end{split}
\end{equation}

\begin{figure}[hbt!]
  \includegraphics[width=.45\textwidth]{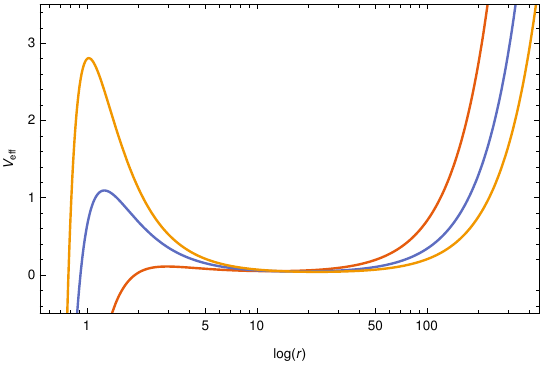}
  \includegraphics[width=.45\textwidth]{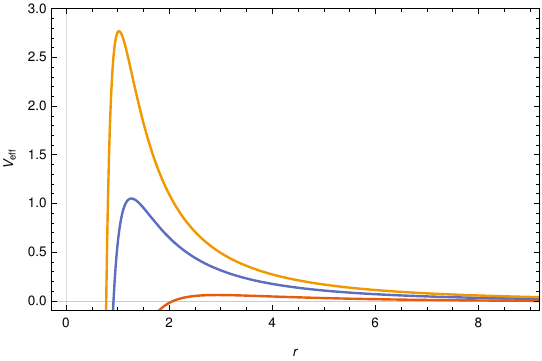}
\caption{Graphs of the effective potential for scalar field for various dimensions: $D=4$ (red), $D=5$ (blue), and $D=6$ (yellow) with LSB parameter $L=2$. The physical parameters are chosen as $M=l=1$ and $\Lambda_e=-0.01$ (left), $\Lambda_e=0.01$ (right).} \label{fig:3a}
\end{figure}

\begin{figure}[hbt!]
  \includegraphics[width=.45\textwidth]{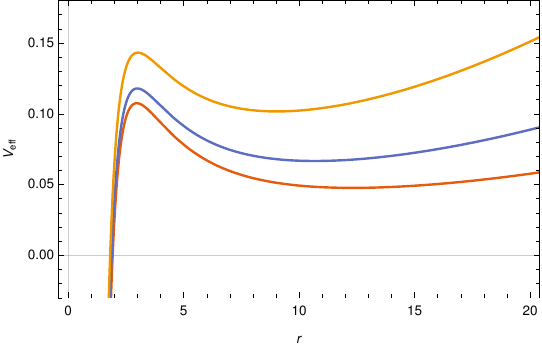}
  \includegraphics[width=.45\textwidth]{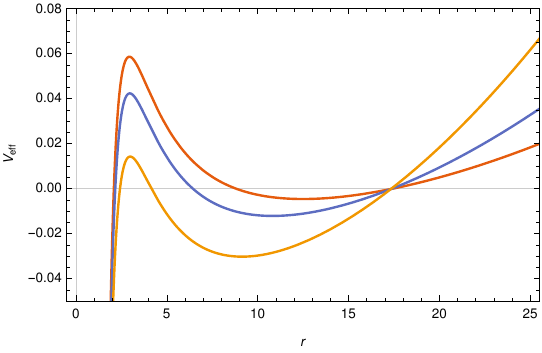}
\caption{Graph of effective potential for scalar particle for various bumblebee parameters $L=2$(red),$L=4$(blue) and $L=8$(yellow) and  dimension $D=4$. The physical parameters are determined as $M=l=1$ and $\Lambda_e=-0.01$(left), $\Lambda_e=0.01$(right).} \label{fig:4a}
\end{figure}

To understand the potential behaviors, we have plotted Eq. \eqref{SS1} for AdS and dS spacetimes in different dimensions (see Fig. \ref{fig:3a}). We have observed that for the AdS spacetime, the potential vanishes only once while for the dS spacetime, it vanishes twice, which is the outcome of the double horizons obtained in the dS spacetime. We have found similar behaviors for the LSB parameter (see Fig. \ref{fig:4a} with fixed $D=4$-dimension for AdS and dS backgrounds).

The general semi-analytic bounds for the GbFs are given by

\begin{equation}
\begin{split}
\sigma(\omega) \geq \sec h^2\left[ \int_{-\infty}^{+\infty} \wp dr_*\right]
\end{split}\label{0S1},
\end{equation}

where

\begin{equation}
\begin{split}
\wp=\frac{\sqrt{(h'^2)+(\omega^2-V_{eff}-h^2)^2}}{2h}
\end{split}. \label{Sara1}
\end{equation}

\begin{figure}[h]
\centering
\includegraphics[scale=.4]{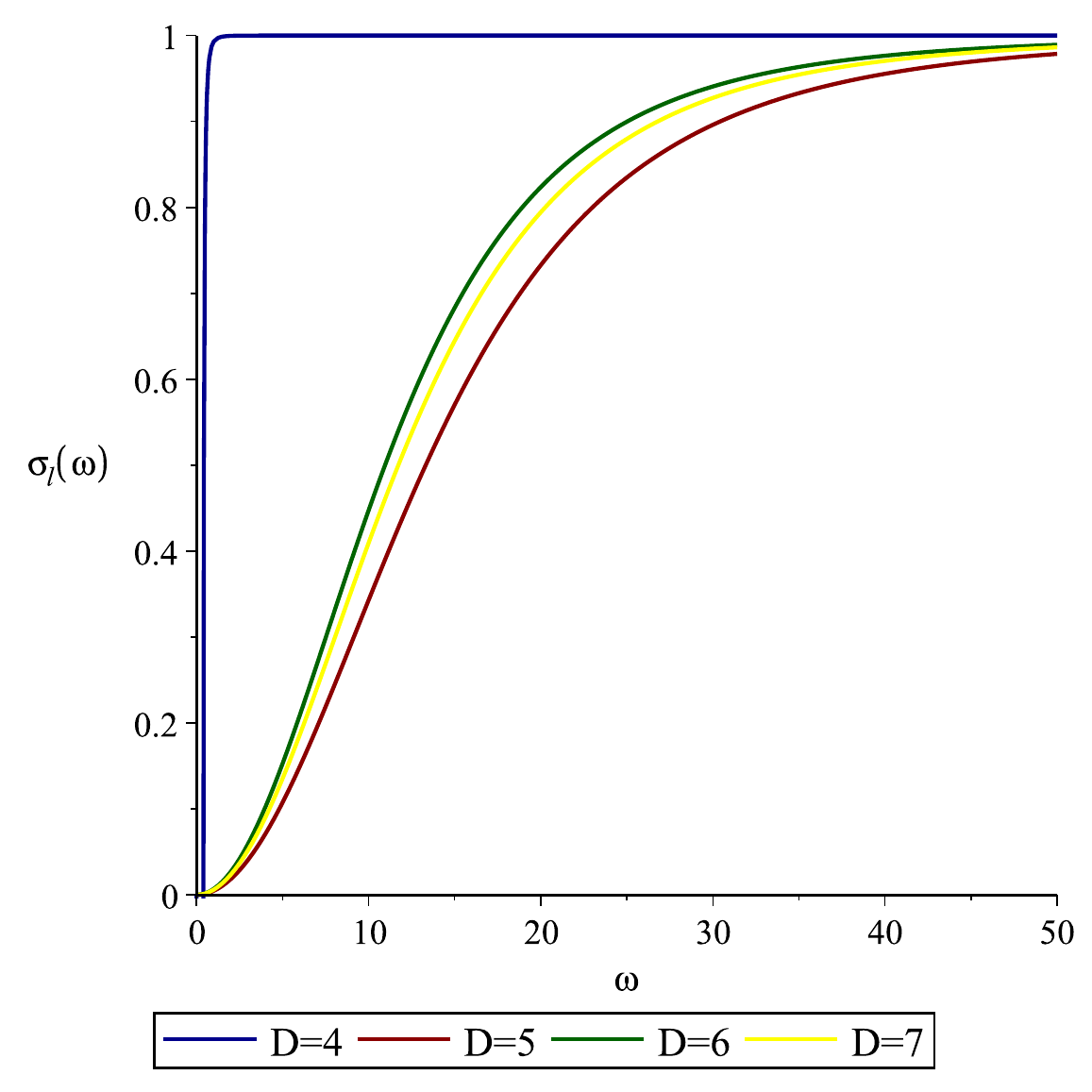} \caption{Graph of scalar GbFs 
for $\Lambda_e<0$ (AdS) and various $D$-dimensions. The physical parameters are chosen as $M=l=1$ and $\Lambda_e=-0.01.$}
\label{FigureS1}%
\end{figure}

We have two conditions for the certain positive function $h: 1) \:h(r_*)>0$ and $2)\: h(-\infty) = h(+ \infty) = \omega $ \cite{sarak1}. After applying the conditions to Eq. \eqref{Sara1}, one may observe a direct proportionality between the GbFs and the effective potential, where the metric function plays a significant part in this process. Since there is no upper border in the integral of Eq. \eqref{0S1}, without loss of generality one can set $h=\sqrt{\omega^2-V_{eff}}$. Thus, Eq. \eqref{0S1} becomes
\begin{equation}
    \sigma_{l}(\omega)\geq sech^2\left[\frac{1}{2}\int_{-\infty}^{+\infty}|\frac{h^{\prime}}{h}|dr_{\ast}\right],\label{S2}
\end{equation}
which results in
\begin{equation}
  \sigma_{l}(\omega)\geq sech^{2}\left[ln(\frac{h_{peak}}{h})\right], \label{S3}  
\end{equation}
where $h_{peak}=\sqrt{\omegat^2-V_{peak}}$. Eq. \eqref{S3} can also be rewritten as 
\begin{equation}
    \sigma_{l}(\omega)\geq\frac{4\omega^2(\omega^2-V_{peak})}{(2\omega^2-V_{peak})}.\label{S4}
\end{equation}
For evaluating $V_{peak}$, first $r_{peak}$ should be determined for different sub-cases. The behaviours of the obtained GbFs for the scalar particles are depicted in Fig. \eqref{FigureS1} for the AdS black hole of the EBGT. What is interesting in Fig. \eqref{FigureS1} is that while the $4$-dimensional black hole has the highest GbF values, the $5$-dimensional black hole has the weakest GbF values. However, the other higher dimensions ($D>5$) have GbF values between the $4_{th}$ and $5_{th}$ dimensions. In $D>5$ dimensional black holes, the GbF values decrease as the dimension increases. Moreover, it is seen that the bosonic GbFs of the AdS bumblebee black hole are almost unaffected by the change in the LSB parameter.

For positive cosmological constant, by considering the second condition, Eq. \eqref{0S1} is expressed by 

\begin{equation}
\begin{split}
\sigma_{l}(\omega) \geq \sec h^2\left[ \frac{\sqrt{1+L}}{2 \omega} \int_{r_h}^{r_H} \frac{V_{eff}}{f(r)} dr\right]
\end{split}, \label{S12}
\end{equation}
whose integration is solvable. Thus, we have\\
\begin{multline}
    \sigma_{l}\geq sech^{2}\left[\frac{\sqrt{1+L}}{2\omega}\left(-\left(\frac{(D-2)(D-4)}{4(1+L)}+l(D+l-3)\right)\frac{1}{r_{H}-r_{h}}-\left(\frac{D\Lambda_e}{2(D-1)}\right)(r_{H}-r_{h})  \right.\right.
\\
\left. \left.
    +\left(\frac{4\pi M(D-4)}{(1+L)\Omega_{D-2}(D-2)}-\frac{8\pi M(D-3)}{(1+L)\Omega_{D-2}(D-2)}\right)\left(\frac{1}{r_{H}^{D-2}}-\frac{1}{r_{h}^{D-2}}\right)
    \right)\right].
\end{multline}

\begin{figure}[h]
\centering
\includegraphics[scale=.4]{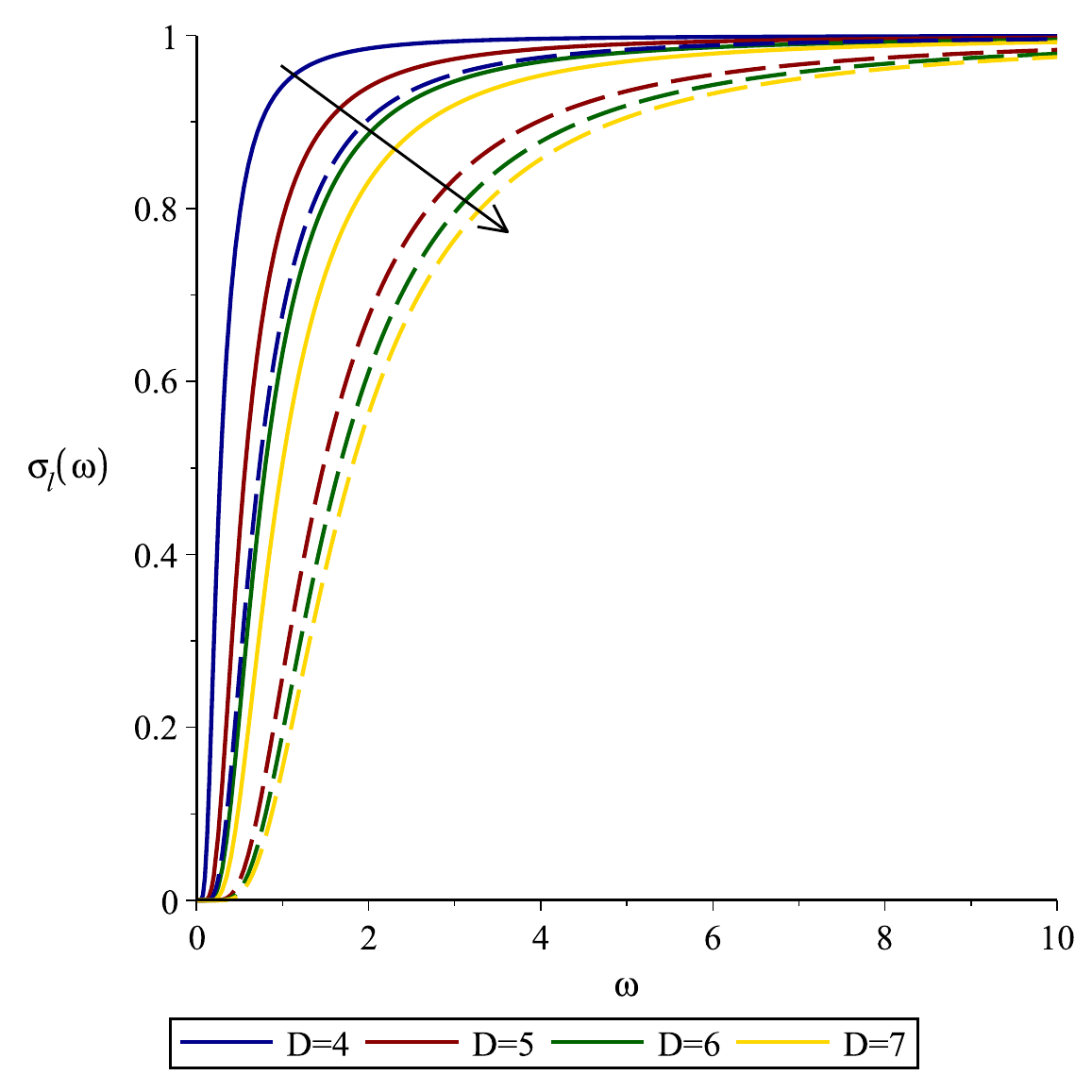} \caption{Graph of scalar GbFs 
for $\Lambda_e>0$ (dS) having various $D$-dimensions and the bumblebee parameter $L$. As the solid lines are for $L=1$, the dashed lines stand for $L=20$. The physical parameter are chosen as $M=l=1$ and $\Lambda_e=0.01$}
\label{FigureS2}%
\end{figure}
The behaviour of GbFs for $\Lambda_e>0$ (dS) is depicted in Fig. \eqref{FigureS2} to show the influences of the dimension and the LSB parameter on the GbFs of the higher dimensional dS black hole in the EBGT. The most important finding from Figs. \eqref{FigureS1} and \eqref{FigureS2} is that the GbF of the 4-dimensional black hole in the EBGT theory is higher than those of its higher dimensional versions. Namely, the GbF drastically reduces with the increasing dimensions, which means that the probability for detecting the thermal radiation of the higher dimensional black holes in the EBGT gets lower with $D>4$.

\section{G\lowercase{b}F\lowercase{s} \lowercase{of} FERMIONS}\label{sec:waveEq}

In this section, we shall investigate the GbFs of the Dirac particles i.e., fermions. To derive the 1-dimensional Schr\"{o}dinger like wave equation, we apply a particular conformal transformation, which contains the Dirac Lagrangian invariant \cite{Cho:2007zi, Chakrabarti:2008xz}. Under the aforementioned conformal transformation \cite{Das:1996we, Gibbons:1993hg}, one has

\begin{equation}
\begin{split}
g_{\mu \nu} \rightarrow \bar{g_{\mu \nu}}=\Omega^2 g_{\mu \nu},
\end{split}
\end{equation}

\begin{equation}
\begin{split}
\psi \rightarrow \bar{\psi} = \Omega^{-(D-1)/2} \psi,
\end{split}
\end{equation}
and
\begin{equation}
\begin{split}
\gamma^\mu \nabla_\mu \psi \rightarrow \bar{\gamma^\mu} \bar{\nabla_\mu} \bar{\psi} = \Omega^{(D+1)/2} \gamma^\mu \nabla_\mu \psi.
\end{split}
\end{equation}

If we consider $\Omega=1/r$, the metric \eqref{25} becomes

\begin{equation}
d\bar{s}^2=-\frac{f(r)}{r^2} dt^2+\frac{1+L}{r^2 f(r)} dr^2 + d\Omega^2_{D-2},
\end{equation}

and

\begin{equation}
\bar{\psi}=r^{(D-1)/2} \psi.
\end{equation}

Thus, the $t-r$ and $(D-2)$-sphere parts of the metric are separated and whence the Dirac equation can be rewritten  as

\begin{equation}
\bar{\gamma^\mu} \bar{\nabla_\mu} \bar{\psi}=0,
\end{equation}
which has the following expansion 
\begin{equation}
[(\bar{\gamma}^t \bar{\nabla}_t + \bar{\gamma}^r \bar{\nabla}_r) \otimes 1]\bar{\psi} + [\bar{\gamma}^5 \otimes (\bar{\gamma}^a \bar{\nabla}_a)_{S_{D-2}}]\bar{\psi}=0, \label{dirac}
\end{equation}

where $(\gamma^5)^2=1$: we can now omit the bar. Furthermore, let us consider $\chi_l^{(\pm)}$ as the eigenspinors for the $(D-2)$-sphere \cite{Camporesi:1995fb}. Then, we have

\begin{equation}
(\gamma^a \nabla_a)_{S_{D-2}} \chi_l^{(\pm)}=\pm i \left(l+\frac{D-2}{2} \right) \chi_l^{(\pm)}, 
\end{equation}

where $l=0,1,2,3...$ . We can also consider $\psi$ as the orthogonal eigenspinors:

\begin{equation}
\psi=\sum_{l} (\phi_l^{(+)} \chi_l^{(+)} + \phi_l^{(-)} \chi_l^{(-)}).
\end{equation}

Therefore, the Dirac equation \eqref{dirac} can be written as follows

\begin{equation}
\left[\gamma^t \nabla_t + \gamma^r \nabla_r + \gamma^5 \left[ \pm i \left( l + \frac{D-2}{2} \right) \right]   \right] \phi_l^{(\pm)}=0,
\end{equation}

where $\gamma^5$ is the interaction term presented in the two dimensional Dirac equation. To tackle with the Dirac equation, we get help from the following auxiliary expressions:

\begin{equation}
\gamma^t=\frac{r}{\sqrt{f(r)}}(-i \sigma^3),
\end{equation}
and
\begin{equation}
\gamma^r=\frac{r \sqrt{f(r)}}{\sqrt{1+L}}( \sigma^2),
\end{equation}

where the $\sigma^i$ are the known Pauli matrices,

\begin{eqnarray}\label{eq6}
\sigma^{1}=\left(
           \begin{array}{cc}
             0 & 1 \\
             1 & 0 \\
           \end{array}
         \right),
 \qquad
         \sigma^{2}=\left(
           \begin{array}{cc}
             0 & -i \\
             i & 0 \\
           \end{array}
         \right),
  \qquad
   \sigma^{3}= \left(
           \begin{array}{cc}
             1 & 0 \\
             0 & -1 \\
           \end{array}
         \right),
\end{eqnarray}

$\gamma^5=(-i\sigma^3)(\sigma^2)=-\sigma^1$. Therefore, we can write the spin connections as

\begin{equation}
\Gamma_t=\sigma^1 \left( \frac{r^2}{4 \sqrt{1+L}} \right) \frac{d}{dr} \left( \frac{f}{r^2} \right).
\end{equation}

We will use the positive sign from now on, as both signs work similarly and we can use either one. Hence, we can explicitly rewrite the Dirac equations as follows

\begin{multline}
    \left[ \frac{r}{\sqrt{f(r)}} (-i \sigma^3) \left[ \frac{\partial}{\partial t} + \sigma^1 \left( \frac{r^2}{4 \sqrt{1+L}} \right) \frac{d}{dr} \left( \frac{f(r)}{r^2} \right) \right] + \frac{r \sqrt{f(r)}}{\sqrt{1+L}} \sigma^2 \frac{\partial}{\partial r} + \right. \\ \left. (-\sigma^1)(i) \left( l+ \frac{D-2}{2} \right) \right] \phi_l^{(+)}=0,
\end{multline}

\begin{multline}
   \sigma^2 \left( \frac{r \sqrt{f(r)}}{\sqrt{1+L}} \right) \left[ \frac{\partial}{\partial r} +\frac{r}{2 \sqrt{f(r)}} \frac{d}{dr} \left( \frac{\sqrt{f(r)}}{r} \right) \right] \phi_l^{(+)}  \\ -i \sigma^1 \left( l+\frac{D-2}{2} \right) \phi_l^{(+)}=i \sigma^3 \left( \frac{r}{\sqrt{f(r)}} \right)\frac{\partial \phi_l^{(+)}}{\partial t}. 
\end{multline}

After letting the following ansatz

\begin{equation}
\begin{split}
\phi_l^{(+)}=\left( \frac{\sqrt{f(r)}}{r} \right)^{-1/2} e^{-i\omega t} \begin{pmatrix}
  iG(r) \\
  F(r) \\
\end{pmatrix},
\end{split}
\end{equation}

we get a simplified form of the Dirac equation:

\begin{equation}
\sigma^2 \left( \frac{r \sqrt{f(r)}}{\sqrt{1+L}} \right) \begin{pmatrix}
  i\frac{dG}{dr} \\
  \frac{dF}{dr} \\
\end{pmatrix} -i \sigma^1 \left( l+ \frac{D-2}{2} \right) \begin{pmatrix}
  iG(r) \\
  F(r) \\
\end{pmatrix}  =i \sigma^3 \omega \left( \frac{r}{\sqrt{f(r)}} \right) \begin{pmatrix}
  iG(r) \\
  F(r) \\
\end{pmatrix}.
\end{equation}
From the above expression, we get

\begin{equation}
\begin{split}
\frac{f(r)}{\sqrt{1+L}} \frac{dG}{dr} - \frac{\sqrt{f(r)}}{r} \left( l+\frac{D-2}{2} \right) G = \omega F, \label{dc1}
\end{split}
\end{equation}

\begin{equation}
\begin{split}
\frac{f(r)}{\sqrt{1+L}} \frac{dF}{dr} + \frac{\sqrt{f(r)}}{r} \left( l+\frac{D-2}{2} \right) G = -\omega G. \label{dc2}
\end{split}
\end{equation}

Recalling the tortoise coordinate $dr_*=\sqrt{1+L}\frac{dr}{f(r)}$ and setting $W=\frac{\sqrt{f(r)}}{r} \left( l+\frac{D-2}{2} \right)$, Eqs. \eqref{dc1} and \eqref{dc2} are simplified to

\begin{equation}
\begin{split}
\left( \frac{d}{dr_*}-W \right)G=\omega F,
\end{split}
\end{equation}

\begin{equation}
\begin{split}
\left( \frac{d}{dr_*}+W \right)F=-\omega G.
\end{split}
\end{equation}
We can now decouple the above equations as
\begin{equation}
\begin{split}
\left(- \frac{d^2}{dr^2_*}+V_1 \right)G=\omega^2 G,
\end{split}
\end{equation}

\begin{equation}
\begin{split}
\left(- \frac{d^2}{dr^2_*}+V_2 \right)F=\omega^2 F,
\end{split}
\end{equation}
where
\begin{equation}
\begin{split}
V_{1,2}=\pm \frac{dW}{dr_*} + W^2.
\end{split}\label{SS2}
\end{equation}

\begin{figure}[hbt!]
  \includegraphics[width=.45\textwidth]{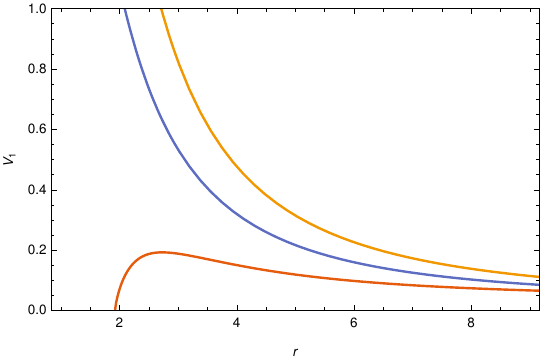}
  \includegraphics[width=.45\textwidth]{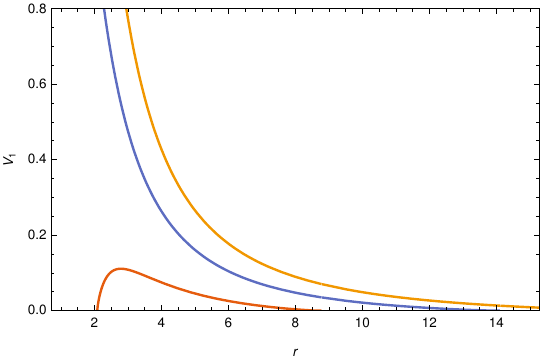}
\caption{Graph of effective potential $V_1$ for Dirac particle for various dimensional values $D=4$ (red), $D=5$ (blue), and $D=6$ (yellow), and $L=2$. The physical parameters are chosen as $M=l=1$ and $\Lambda_e=-0.01$ (left), $\Lambda_e=0.01$ (right).} \label{fig:7a}
\end{figure}

\begin{figure}[hbt!]
  \includegraphics[width=.45\textwidth]{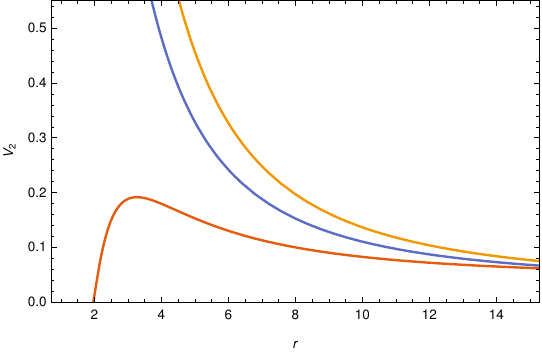}
  \includegraphics[width=.45\textwidth]{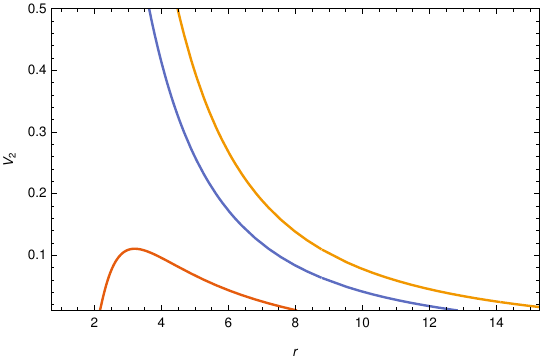}
\caption{Graph of effective potential $V_2$ for Dirac anti-particle for various dimensional values $D=4$ (red), $D=5$ (blue), and $D=6$ (yellow), and $L=2$. The physical parameters are chosen as $M=l=1$ and $\Lambda_e=-0.01$ (left), $\Lambda_e=0.01$ (right).} \label{fig:8a}
\end{figure}

These two potentials $V_{1,2}$ belong to the particle and anti-particles of Dirac fermions. We have shown the behaviour of the potentials  $V_{1,2}$ for the different dimensions in dS and AdS space (see Figs. \ref{fig:7a} and \ref{fig:8a}). Depending on the existence of the cosmological horizon in AdS/dS spacetimes, the potential vanishes at some radial distance as being observed in the scalar potential.

Now, by considering the above potentials \eqref{SS2} in Eq. \eqref{S12}, the fermionic GbFs of the EBGT, for various dimensions, are obtained as the following: \\
For $D=4$:
\begin{equation}
    \sigma_{l,D=4}(\omega)\geq sech^{2}\left[\frac{1}{2\omega}\left(\pm\frac{3(l+1)}{\sqrt{-3\gamma}}\left[\frac{M}{r_{h}^{3}}\left(1+\frac{9}{10\gamma r_{h}^2}\right)-\frac{1}{2r_{h}^2}(1+\frac{3}{4\gamma r_{h}^2})\right]+
    \frac{\sqrt{1+L}(1+l)^2}{r_{h}}\right)\right],\label{SK2}
\end{equation}
for $D=5$:
\begin{equation}
    \sigma_{l,D=5}(\omega)\geq sech^2\left[\frac{1}{2\omega}\left(\pm\frac{6(l+\frac{3}{2})}{\sqrt{-6\gamma}}\left[\frac{M}{\pi r_{h}^4}\left(1+\frac{4}{3\gamma r_{h}^2}\right)- \frac{1}{2r_{h}^2}(1+\frac{3}{2\gamma r_{h}^2})\right]+
    \frac{\sqrt{1+L}(l+\frac{3}{2})^2}{r_{h}}\right)\right], \label{SK3}
\end{equation}

and for $D=6$:
\begin{equation}
    \sigma_{l,D=6}(\omega)\geq sech^2\left[\frac{1}{2\omega}\left(\pm\frac{(l+2)}{\sqrt{-0.1\gamma}}\left[\frac{M}{\pi r_{h}^5}\left(\frac{50}{18.62\gamma r_{h}^2}+0.752\right)-\frac{1}{2r_{h}}(1+\frac{5}{2\gamma r_{h}^2})\right]+
   \frac{\sqrt{1+L}(l+2)^2}{r_{h}} \right)\right]. \label{SK4}
\end{equation}
In the above results [Eqs. \eqref{SK2}-\eqref{SK4}], $\gamma=\Lambda_e (1+L)$, the asymptotic series approach is utilized in order to facilitate the integration evaluations. That is why the GbFs are served in discrete forms for different dimensions.\\
Since the integration is bounded between the outer and cosmological horizons, to define the GbFs of the dS black hole, we directly evaluate the integration \eqref{S12} and get

\begin{multline}
    \sigma_{l}(\omega)\geq sech^2\left[\frac{1}{2\omega}\left(\pm(l+\frac{D-2}{2})(\frac{1}{r_{H}}-\frac{1}{r_{h}})\sqrt{1-\frac{16\pi M}{(D-2)\Omega_{D-2}}(\frac{1}{r_{H}^{D-3}}-\frac{1}{r_{h}^{D-3}})\frac{2(1+L)\Lambda_e (r_{H}^2-r_{h}^2)}{(D-1)(D-2)}}\right.\right.
    \\
    \left.\left.
    +\sqrt{1+L}(l+\frac{D-2}{2})^2(\frac{1}{r_{H}}-\frac{1}{r_{h}})
    \right)\right],\label{SK5}
\end{multline}

\begin{figure}[hbt!]
  \includegraphics[width=.45\textwidth]{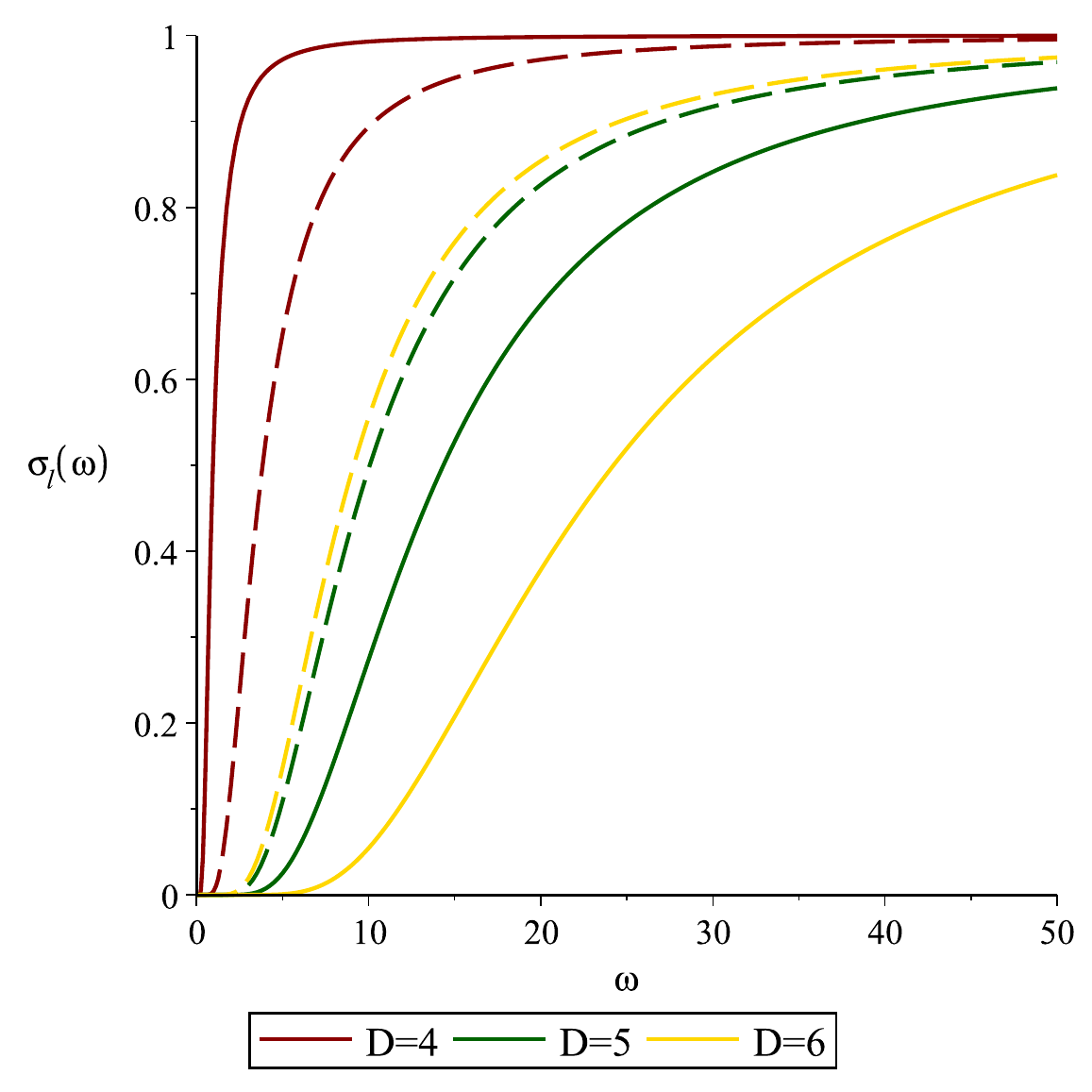}
  \includegraphics[width=.45\textwidth]{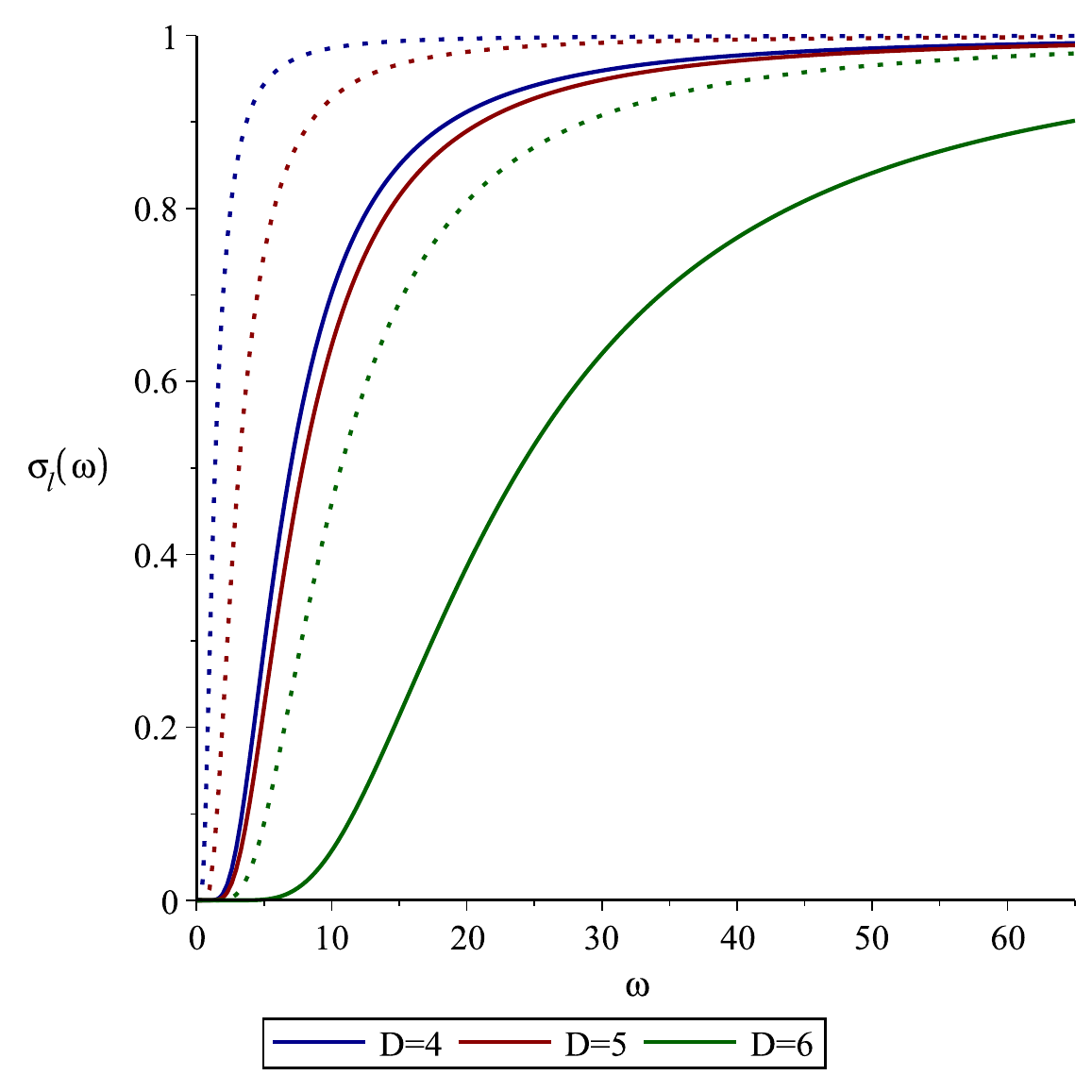}
\caption{Graph of Dirac GbFs for $\Lambda_e=-0.1$ (left) and $\Lambda_e=0.1$ (right) in various dimensions $D\geq4$. As the solid lines represent GbFs for spin-up, the dashed and dotted lines stand for the GbFs of the spin-down particles. The physical parameters are chosen as $M=l=1$.} \label{fig:S8a}
\end{figure}




The behaviors of the Dirac GbFs for various dimensions and LSB parameters are illustrated in Fig. \eqref{fig:S8a} for negative (left) and positive (right) cosmological constants. It can be deduced from the associated figures that the most highest fermionic GbFs in both cases (dS/AdS) belong to the $D=4$ case. The increase in the dimension decreases the GbFs. In order to have a wider perspective also on the impression of parameter L in the GFs of the bumblebee model refer to \cite{sarak1} as an example of constant dimension $D=4$.

\begin{table}
   \centering
   \begin{tabular}{|c|c|c|c|c|c|c|c|}
\hline
$L$ & $l$ & $n$ & $D$ & $\omega_{Bosons}$ & $L$ & $\omega_{Bosons}$\\
\hline
1 & 0 & 0 & 4 & 0.142524815-0.0271009441i & 1.5 &
0.154034389-0.0293764932i\\
&  &  & 6 & 1.224706041-0.7060164222i &  &
1.313934061-0.6188707214i\\
&  &  & 7 & 1.528815173-1.120609758i &  &
1.427602066-1.283657004i\\
&  &  & 8 & 1.750953184-1.412550460i &  &
1.580613654-1.602390607i\\
\hline
1 & 1 & 0 & 4 & 0.0525299867-0.0191011204i & 1.5 &
0.0507239756-0.0219589547i\\
&  &  & 6 & 7.008944298-2.702402326i &  &
9.515635233-2.793914762i\\
&  &  & 7 & 4.436000670-5.493917855i &  &
5.015834961-7.581928057i\\
&  &  & 8 & 3.547057113-5.730052528i &  &
3.718946501-7.713072092i\\
\hline
\end{tabular}
  \caption{Bosonic QNMs of various dimensional
dS/AdS black holes}, \label{t1}
\end{table}

\begin{figure}[hbt!]
  \includegraphics[width=.45\textwidth]{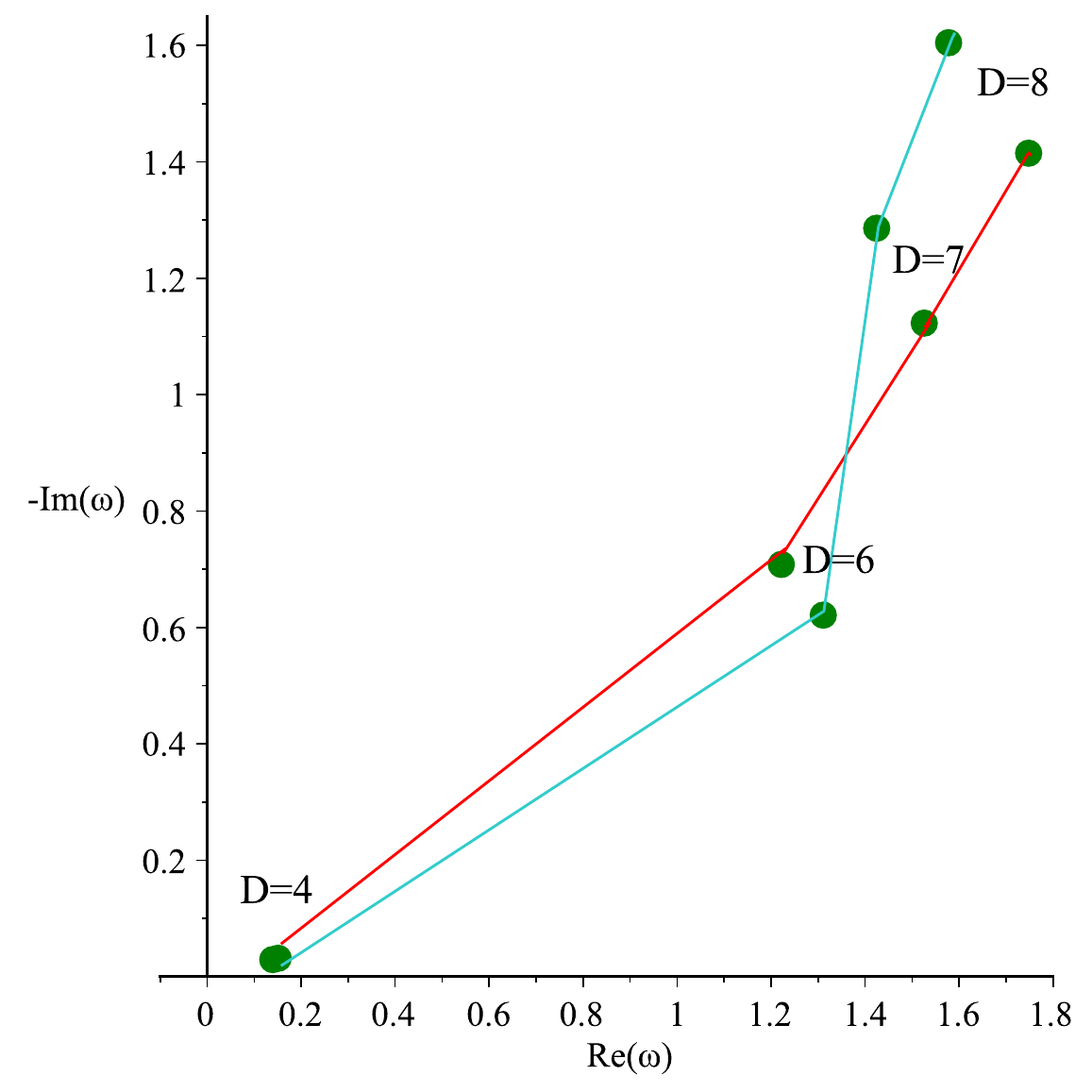}
  \includegraphics[width=.45\textwidth]{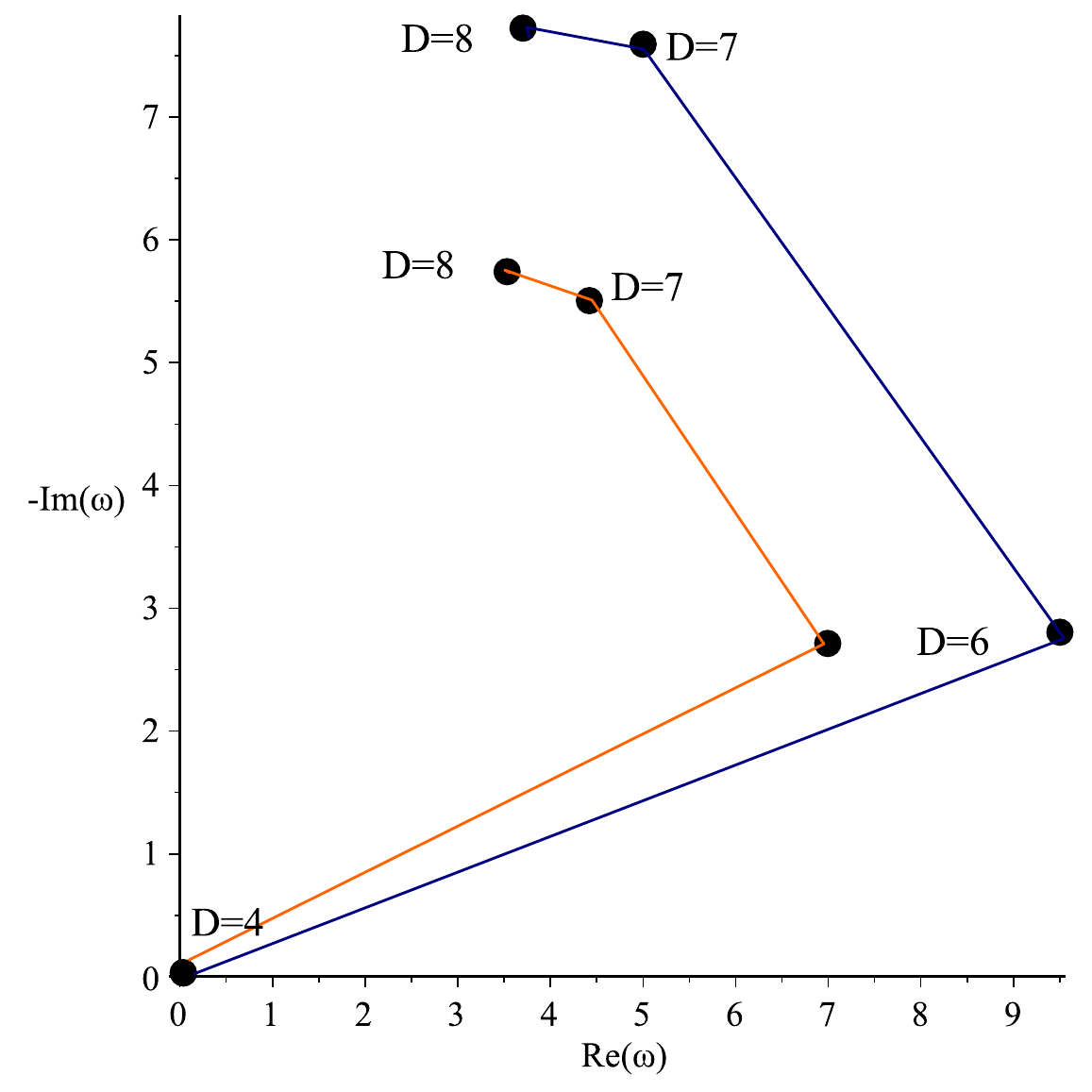}
\caption{Graph of QNMs for the scalar particle for various dimensions; Left figure stands for $l=0$ in which the green line represents $L=1$ and red line is for $L=1.5$. Right figure stands for $l=1$ in which the orange line is for $L=1$ and blue line exhibits $L=1.5$. The physical parameters are chosen as $M=1$ and $\Lambda_e=-0.01$.} \label{fig:SK8}
\end{figure}

\section{QNM\lowercase{s}}\label{sec:QNMs}
The WKB (Wentzel, Hendrik Kramers and Léon Brillouin) approach is an approximate technique to solve the linear differential equations.
The most significant utilization of the WKB approximation is to solve the time independent Schr\"{o}dinger equation.
In general, the equations conducting different types of non-rotating or static black hole QNM perturbations form in terms of the radial coordinate: see Eq. \eqref{sk11}, in which $\omega$ stands for the complex QNMs.\\

\begin{table}[H]
   \centering
   \begin{tabular}{|c|c|c|c|c|c|c|c|}
\hline
$L$ & $l$ & $n$ & $D$ & $\omega_{Fermionic}$ & $L$ & $\omega_{Fermionic}$\\
\hline
1 & 0 & 0 & 4 & 0.0458857289-0.018095089i & 1.2 &
0.109578578+0.0676419223i\\
&  &  & 6 & 0.1242495317-0.8544642331i &  &
0.6694131812-0.7304785062i\\
&  &  & 7 & 2.732511256-2.545148685i &  &
2.832991755-2.928508941i\\
&  &  & 8 & 3.973248774-1.565370350i &  &
4.440682841-1.442269634i\\
\hline
1 & 1 & 0 & 4 & 0.0406050117-0.0158503234i & 1.2 &
0.0922648051+0.0559755816i\\
&  &  & 6 & 0.1111233295-0.7679995927i &  &
0.6038037915-0.6574441601i\\
&  &  & 7 & 5.228007931-5.923490838i &  &
5.525117190-6.784020908i\\
&  &  & 8 & 8.055358919-2.285136166i &  &
8.678911730-2.016127470i\\
\hline
\end{tabular}
  \caption{Fermionic QNMs of various dimensional
dS/AdS black holes}, \label{t2}
\end{table}

\begin{figure}[hbt!]
  \includegraphics[width=.45\textwidth]{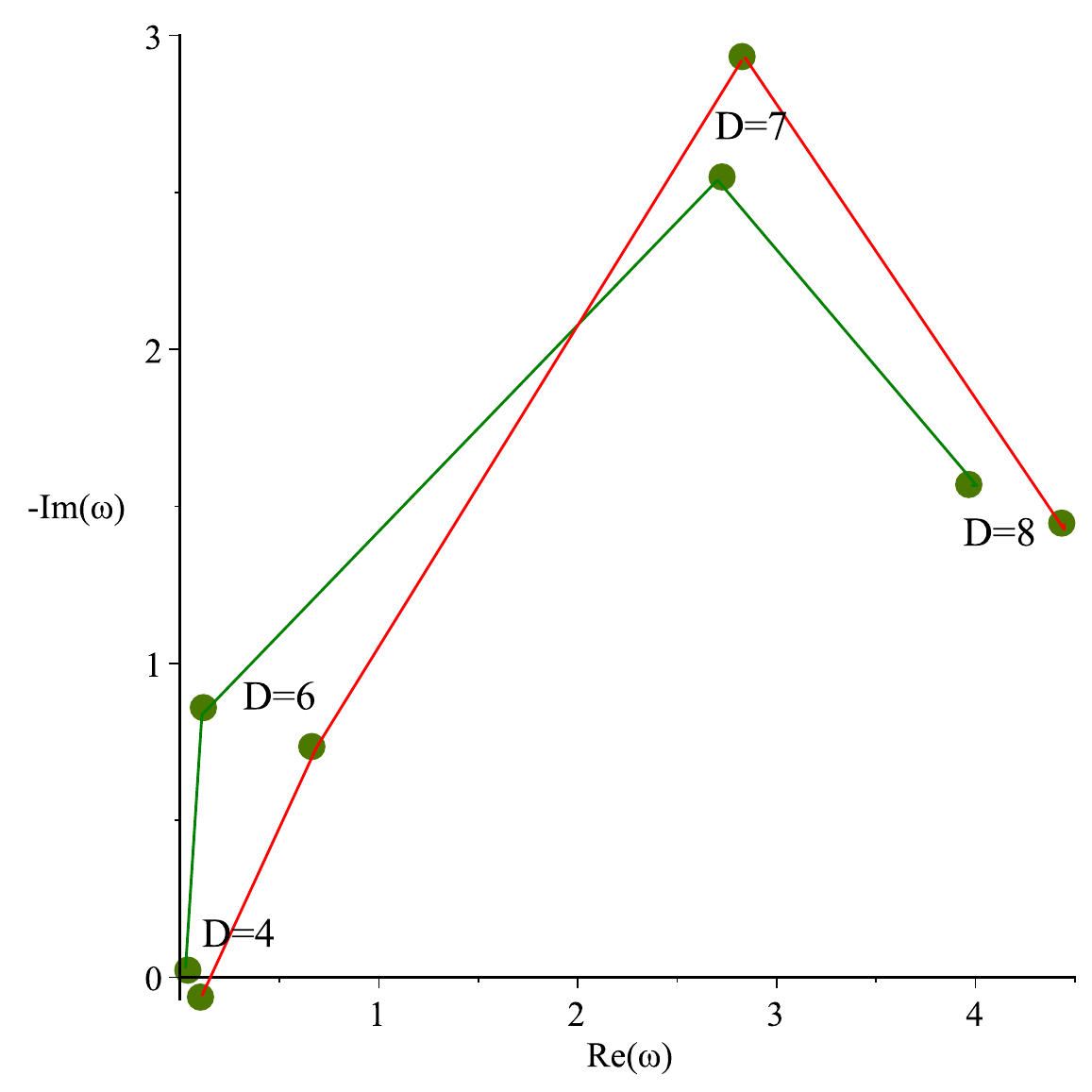}
  \includegraphics[width=.45\textwidth]{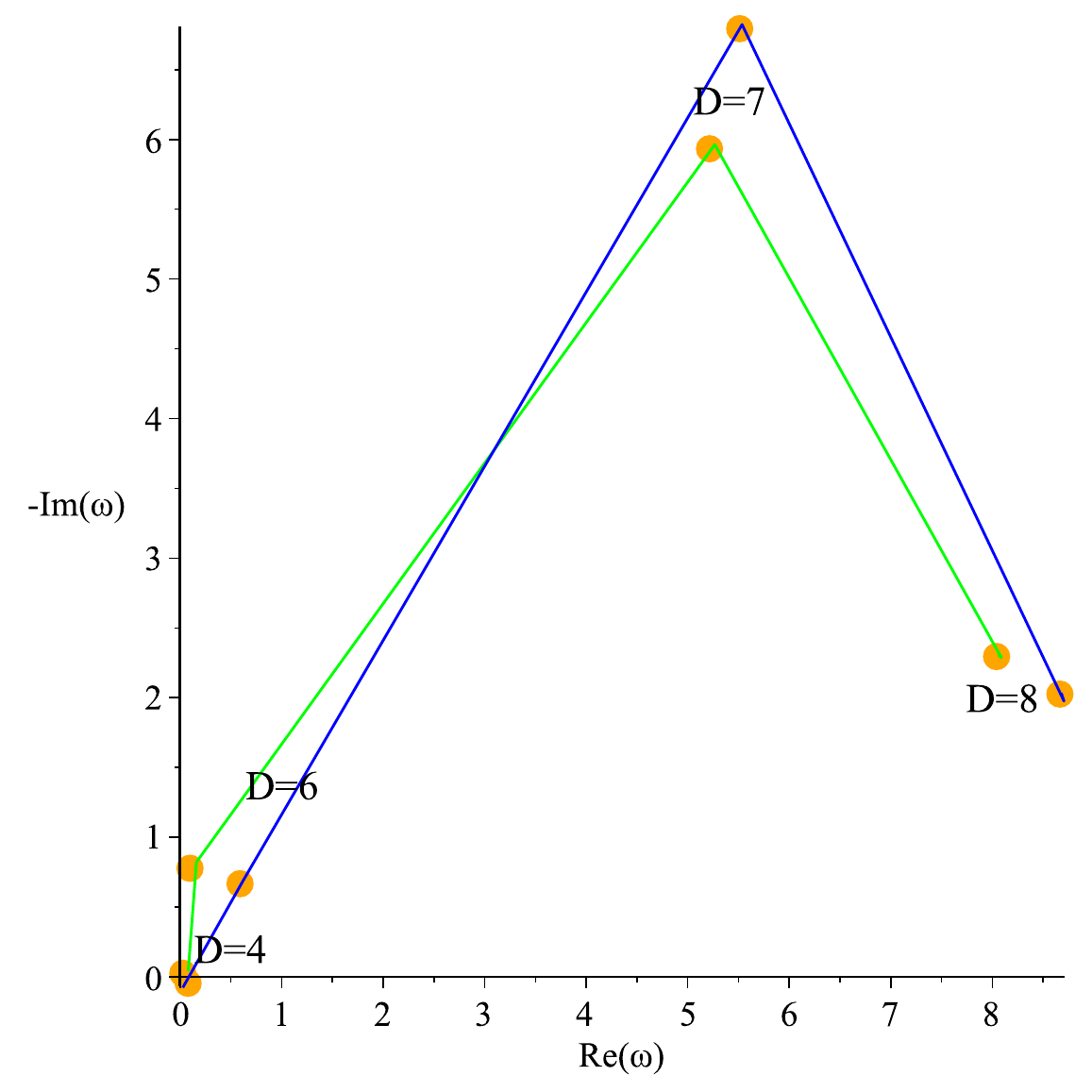}
\caption{Graph of QNMs for a Dirac particle for various dimensions; The left-hand side figure stands for $l=0$ in which the green line represents $L=1$ and red line is for $L=1.2$. The right-hand side figure stands for $l=1$ in which the green line represents $L=1$ and blue line is for $L=1.2$. The physical parameters are chosen as $M=1$ and $\Lambda_e=-0.05$.} \label{fig:S8}
\end{figure}

As is well-known, in quantum mechanics, the $\omega^{2}$ parameter corresponds to $\frac{2m}{\bar{h}^{2}}E$, where $E$ indicates the particle energy. In this context, the effective potential is nothing but a barrier. In order to compute the QNMs, the appropriate boundary conditions must be considered at $(r_{\ast}\rightarrow-\infty)$ and $(r_{\ast}\rightarrow\infty)$, which stand for the event horizon and spatial infinity, respectively. The ingoing modes represent waves moving away from the potential barrier. Namely, the ingoing waves $(r_{\ast}\rightarrow-\infty)$ correspond to the radiation which crosses the horizon into the black hole. On the other hand, since a QNM is occurred from a black hole's free oscillation, the ingoing modes at spatial infinity are ignored. In other words, only outgoing waves exist at spatial infinity, $(r_{\ast}\rightarrow\infty)$. \\

To compute the QNMs with the WKB approach, we employ the following complex frequency expression \cite{sarak2}

\begin{equation}
 \omega^{2}=\left[V_{0}+\sqrt{-2V_{0}^{\prime\prime}}\Lambda(n)-i(n+\frac{1}{2})\sqrt{-2V_{0}^{\prime\prime}}(1+\Omega(n))\right], \label{s15}  
\end{equation}
by which 
\begin{equation}
    \Lambda(n)=\frac{1}{\sqrt{-2V_{0}}^{\prime\prime}}\left[\frac{1}{8}(\frac{V_{0}^{(4)}}{V_{0}^{\prime\prime}})(\frac{1}{4}+\alpha^2)-\frac{1}{288}(\frac{V_{0}^{\prime\prime\prime}}{V_{0}^{\prime\prime}})^{2}(7+60\alpha^2)\right], \label{s16}
\end{equation}

and
\begin{multline}
\Omega\left(  n\right)  =\frac{1}{-2V_{0}^{\prime\prime}}\left[  \frac{5}{6912}\left(  \frac{V_{0}^{\prime\prime\prime}}{V_{0}^{\prime\prime
}}\right)  ^{4}\left(  77+188\alpha^{2}\right)  -\frac{1}{384}\left(
\frac{V_{0}^{\prime\prime\prime 2}V_{0}^{\left(  4\right)  }}{V_{0}^{\prime\prime3}}\right)  \left(  51+100\alpha^{2}\right)  +\right.  \label{28}\\
\left.  \frac{1}{2304}\left(  \frac{V_{0}^{\left(  4\right)  }}{V_{0}^{\prime\prime}}\right)  ^{2}\left(  67+68\alpha^{2}\right)  +\frac{1}{288}\left(
\frac{V_{0}^{\prime\prime\prime}V_{0}^{\left(  5\right)  }}{V_{0}^{\prime\prime 2}}\right)  \left(  19+28\alpha^{2}\right)  -\frac{1}{288}\left(  \frac
{V_{0}^{\left(  6\right)  }}{V_{0}^{\prime\prime
}}\right)  \left(  5+4\alpha^{2}\right)  \right]  ,
\end{multline}

where the primes and superscript (n = 4, 5, 6; for the higher order derivatives) denote the differentiation with respect to $r_{\ast}$. Furthermore, the subscript $0$ represent the maximum point for the potential. To derive the QNMs of scalar and Dirac particles, we have used the effective potential expressions which are represented in  Eqs. \eqref{SS1} and \eqref{SS2}, respectively. The results of the $6^{th}$ order WKB approach for bosonic particles QNMs are tabulated in Table \eqref{t1}, which reveals that both frequency and damping modes for bosonic QNMs rise (when $l=0$) by increasing the dimension $D$ and/or the LSB parameter $L$.  Contrarily, the picture alters after $D\geq6$. Table \eqref{t1} also shows that for $l=1$ the damping rate grows by rising both $D$ and $L$ parameters, but the real part fluctuates. In more concise evaluation, the results tabulated in Table \eqref{t1} are depicted by Fig. \eqref{fig:SK8}. \\

Table \eqref{t2} represents the fermionic QNMs for $l=0,1$ for various dimensionalities and the LSB parameters. As it is shown in Table \eqref{t2}, both real and imaginary parts of the fermionic QNMs are growing while the dimension increases. Almost the same behaviors are obtained for a rising $L$ parameter except for the damping mode obtained at $D=8$. The behaviors of the associated QNMs are shown in Fig. \eqref{fig:S8}. It is worth mentioning that the bosonic QNMs for $\Lambda>0$ yield almost the same results as those for $\Lambda<0$. In the case of fermionic QNMs, we have observed that the real terms for $\Lambda<0$ correspond to the imaginary terms for $\Lambda>0$. This implies that both positive and negative values of $\Lambda$ exhibit similar behavior.

\section{Null Geodesics and Shadow Radius}\label{sec:Shadow Radius}

In this section, we shall study the photon's orbit and radius of the shadow of the black hole \cite{Gullu:2020qzu}. Let us first consider the Lagrangian $\mathcal{L}(x,\Dot{x})=(1/2)g_{\mu \nu} \Dot{x}^\mu \Dot{x}^\nu$ for the static spherically symmetric metric, which is given by \eqref{25}

\begin{equation}
    \mathcal{L}(x,\Dot{x})=\frac{1}{2} \left( -f(r) \Dot{t}^2+\frac{(1+L)}{f(r)} \Dot{r}^2 + r^2 d \Omega^2_{D-2} \right).
\end{equation}

The spacetime has two conserved quantities which can be calculated by solving the Euler-Lagrange equations in the equatorial plane. Hence, the conserved quantities are obtained as 
\begin{equation}
    \begin{split}
        E=f(r)\Dot{t}, \qquad \tilde{L}=r^2\Dot{\phi},
    \end{split}
\end{equation}

where $E$ and $\tilde{L}$ are called the conserved specific energy and conserved specific angular momentum. For the photons, we can write the following equation:

\begin{equation}
    0=-f(r) \Dot{t}^2+\frac{(1+L)}{f(r)} \Dot{r}^2 + r^2 d \Omega^2_{D-2},
\end{equation}

which gives the following equation by the aid of the conserved quantities

\begin{equation}\label{75}
    \left( \frac{dr}{d\phi} \right)^2 = V_{eff}, 
\end{equation}

where

\begin{equation}
    V_{eff}=\frac{r^2 f(r)}{(1+L)} \left(  \frac{r^2}{f(r)} \frac{E^2}{\tilde{L}^2}-1 \right ).
\end{equation}

Now we define the impact parameter $b=\tilde{L}/E$. At the photon sphere radius $r=r_{ph}$, the conditions $dr/d\phi |_{r_{ph}}=0$ ($V_{eff}=0$) and $V'_{eff}=0$ should be satisfied. Hence, the impact parameter for the photon sphere is given by

\begin{equation}
    \frac{1}{b^2}=\frac{f(r_{ph})}{r^2_{ph}}.
\end{equation}

By using conditions $dr/d\phi |_{r_{ph}}=0$ and $d^2r/d\phi^2 |_{r_{ph}}=0$ \cite{Luminet:1979nyg}, we can find the radius of the photon sphere as

\begin{equation}
    \frac{d}{dr} B(r)^2=0,
\end{equation}

where $B(r)=\sqrt{\frac{r^2}{f(r)}}$. Therefore, Eq. \eqref{75} becomes

\begin{equation}
    \left( \frac{dr}{d\phi} \right)^2 = \frac{r^2 f(r)}{(1+L)} \left(  \frac{B^2(r)}{B^2(r_{ph})} -1 \right ).
\end{equation}

Now, we define an angle $\alpha$ \cite{Perlick:2021aok} between the null light ray and radial direction as follows

\begin{equation}
    \cot{\alpha} = \frac{\sqrt{1+L}}{\sqrt{r^2 f(r)}} \frac{dr}{d\phi} \Bigr|_{r=r_0},
\end{equation}
which gives
\begin{equation}
    \cot^2{\alpha}=\frac{B^2(r_0)}{B^2(r_{ph})}-1.
\end{equation}

Using some trigonometric relations, one can get the following equation:

\begin{equation}
    \sin^2{\alpha}=\frac{B^2(r_{ph})}{B^2(r_0)}.
\end{equation}

We define $b_{cr}$ as the impact parameter at the critical impact parameter, thus we have

\begin{equation}
    \sin^2{\alpha}_{sh}=\frac{b^2_{cr}}{B^2(r_0)}.
\end{equation}

Ultimately, the radius of the shadow for the static observer at $r=r_0$ \cite{Konoplya:2019sns} is found to be

\begin{equation}
    R_{sh}=r_0 \sin{\alpha}=\sqrt{\frac{r^2_{ph} f(r_0)}{f(r_{ph})}},
\end{equation}


and by considering a mathematical constraint $f(r_0) \approx 1$ for a static observer located at a special location (see for example \cite{Sakalli:2014bfa,Mirekhtiary:2019ugi} in which the appropriate normalization for the time-like Killing vector was applied), we can write the shadow radius \cite{Du:2022hom} as follows

\begin{equation}
    R_{sh}=\sqrt{\frac{r^2_{ph}}{f(r_{ph})}}.
\end{equation}

Now, in terms of the celestial coordinate, the shadow radius is obtained as

\begin{equation}\label{86}
    X=\lim_{r_0 \to \infty} \left( -r^2_0 \sin{\theta_0} \frac{d\phi}{dr} \Bigr|_{(r_0,\theta_0)}  \right ), \qquad
    Y=\lim_{r_0 \to \infty} \left( r^2_0  \frac{d\theta}{dr} \Bigr|_{(r_0,\theta_0)}  \right ),
\end{equation}

where $(r_0,\theta_0)$ is the position of the observer at spatial infinity. Moreover, since we analyze the shadow of the black hole in the equatorial plane, the shadow of the radius is equivalent to the critical impact parameter of the photon sphere. Therefore, we have

\begin{equation}
    R_{sh}=\sqrt{X^2+Y^2}=b_{cr},
\end{equation}

which is explicitly written as

\begin{equation}\label{92}
    \begin{split}
      R_{sh} =  \frac{(8 \pi )^{-\frac{1}{3-D}} \left(\frac{(D-2) \Omega _{D-2}}{D-1}\right){}^{\frac{1}{3-D}}}{\sqrt{-\frac{2^{1-\frac{6}{3-D}} \pi ^{-\frac{2}{3-D}} \Lambda_e  (L+1) \left(\frac{(D-2) \Omega _{D-2}}{D-1}\right){}^{\frac{2}{3-D}}}{(D-2) (D-1)}-\frac{16 \pi  \left((8 \pi )^{-\frac{1}{3-D}} \left(\frac{(D-2) \Omega _{D-2}}{D-1}\right){}^{\frac{1}{3-D}}\right)^{3-D}}{(D-2) \Omega _{D-2}}+1}}.
    \end{split}
\end{equation}

\begin{figure}[h]
\centering
\includegraphics[width=.45\textwidth]{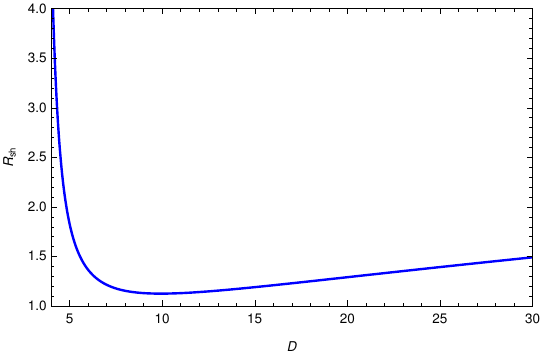}
\includegraphics[width=.44\textwidth]{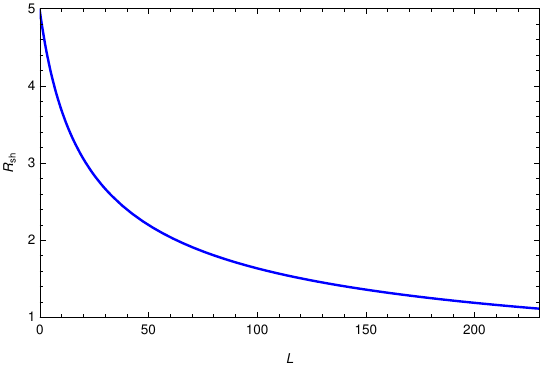}
\caption{In the left-hand side we have shown the shadow radius of the bumblebee black hole versus $D$-dimension. The physical parameters are chosen as $M=L=1$ and $\Lambda_e=-0.01$. On the right-hand side, we have shown the shadow radius of the bumblebee black hole versus the bumblebee parameter $L$. The physical parameters are chosen as $M=1$, $D=4$, and $\Lambda_e=-0.01$.}
\label{sh1}
\end{figure}

\begin{figure}[h]
\centering
\includegraphics[width=.45\textwidth]{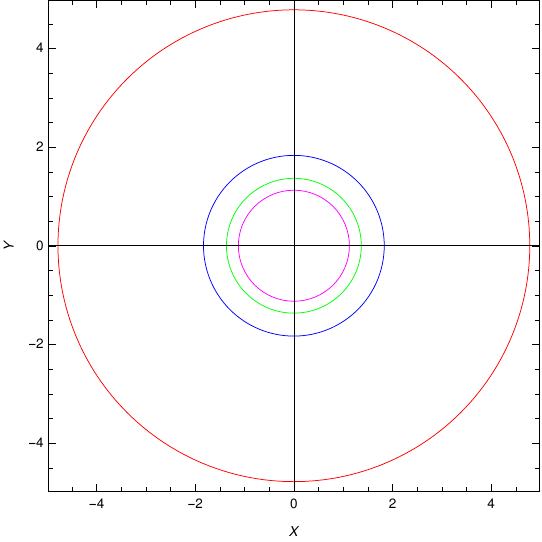}
\includegraphics[width=.45\textwidth]{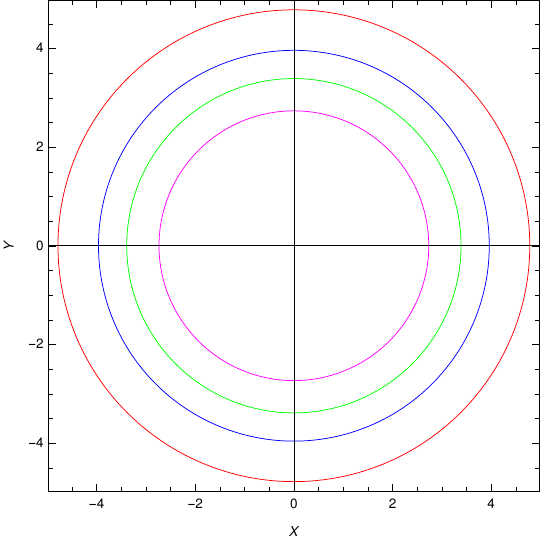}
\caption{2D plot of the shadow in the celestial coordinate $(X,Y)$. Plots are governed by Eq. \eqref{86}. On the left-hand side, the physical parameters are chosen as $M=L=1$ and $\Lambda_e=-0.01$. Each color represents different space-time dimensions: Red ($D=4$), blue ($D=5$), green ($D=6$), and magenta ($D=10$). On the right-hand side, the physical parameters are chosen as $M=1$, $D=4$, and $\Lambda_e=-0.01$. Each color represents different bumblebee parameter: Red ($L=1$), blue ($L=7$), green ($L=14$), and magenta ($L=28$).}
\label{sh2}
\end{figure}

We have shown the variation of the shadow radius of the black hole with the space-time dimension in Fig. \ref{sh1} (left) for $\Lambda_e=-0.01$. After initially declining, the shadow radius then begins to rise. A similar effect is observed in the celestial coordinate, which is shown in Fig. \ref{sh2} (left). Similarly, we have shown the variation of the shadow radius with the bumblebee parameter ($L$) in Fig. \ref{sh1} (right) for $D=4$ and $\Lambda_e=-0.01$. The shadow radius in this case keeps on decreasing with $L$ and the shadow of the black hole is depicted in celestial coordinate in Fig. \ref{sh2} (right).

\begin{figure}[h]
\centering
\includegraphics[width=.45\textwidth]{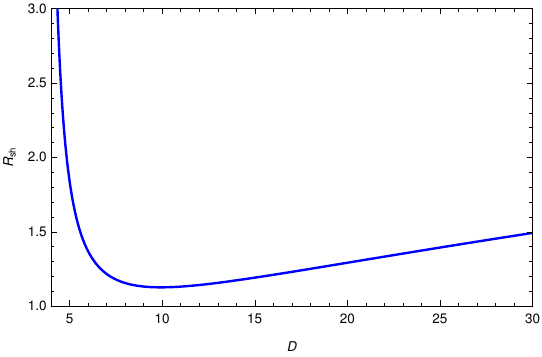}
\includegraphics[width=.44\textwidth]{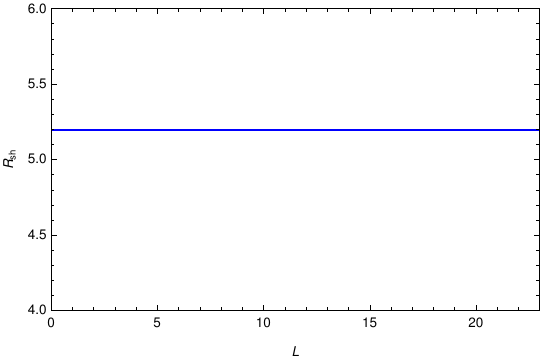}
\caption{In the left-hand side we have shown the shadow radius of the bumblebee black hole versus $D$-dimension. The physical parameters are chosen as $M=L=1$ and $\Lambda_e=1.11*10^{-52} m^{-2}$. On the right-hand side, we have shown the shadow radius of the bumblebee black hole versus the bumblebee parameter $L$. The physical parameters are chosen as $M=1$, $D=4$, and $\Lambda_e=1.11*10^{-52} m^{-2}$.}
\label{sh3}
\end{figure}

\begin{figure}[h]
\centering
\includegraphics[width=.45\textwidth]{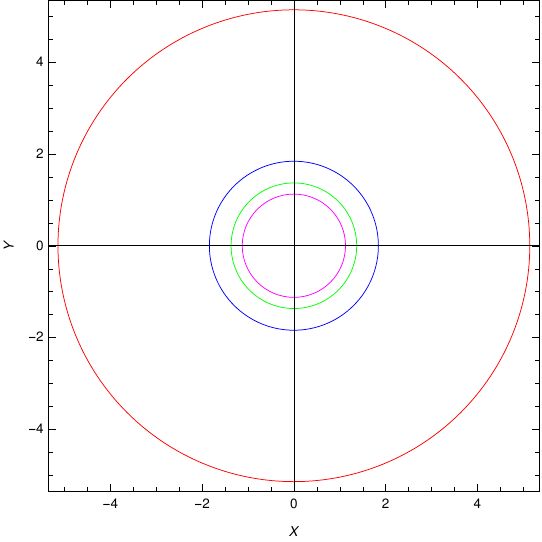}
\includegraphics[width=.45\textwidth]{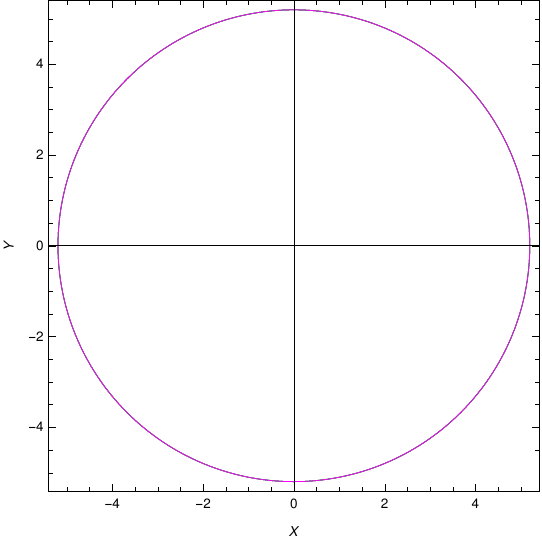}
\caption{2D plot of the shadow in the celestial coordinate $(X,Y)$. Plots are governed by Eq. \eqref{86}. On the left-hand side, the physical parameters are chosen as $M=L=1$ and $\Lambda_e=1.11*10^{-52} m^{-2}$. Each color represents different space-time dimensions: Red ($D=4$), blue ($D=5$), green ($D=6$), and magenta ($D=10$). On the right-hand side, the physical parameters are chosen as $M=1$, $D=4$, and $\Lambda_e=1.11*10^{-52} m^{-2}$. Since the shadow size does not depend on the bumblebee parameter, we have the same shadow size for all $L$.}
\label{sh4}
\end{figure}

We also looked at the shadow size for the dS space and took the observed cosmological constant value $\Lambda_e=1.11*10^{-52} m^{-2}$. We first showed the variation of the shadow size with space-time dimension $D$ in Fig. \ref{sh3} (left) for $M=L=1$. The shadow size variation is similar to the case of Ads but with a larger radius. A similar effect is observed in the celestial coordinate, which is shown in Fig. \ref{sh4} (left). We also observed that the bumblebee gravity parameter does not affect the shadow size at all in any dimension and we particularly have plotted the shadow size for $D=4$ case in Fig. \ref{sh3} (right) for $M=1$ and the shadow of the black hole is depicted in celestial coordinate in Fig. \ref{sh4} (right).

\section{Relationship between Shadow Radius and QNM\lowercase{s}}\label{sec:Relations}

In this section, we will try to reveal the relation between the shadow radius and QNMs. As shown in \cite{Stefanov:2010xz,Jusufi:2019ltj,Cuadros-Melgar:2020kqn,Moura:2021eln,Liu:2020ola}, the real part of the QNMs at the eikonal limit corresponds to the angular velocity of the critical null circular orbit $\Omega_c$ and the imaginary part of the QNMs is nothing but the Lyapunov exponent $\lambda$, which is used to determine the unstable timescale of the circular orbit \cite{Cardoso:2008bp}. Namely, we have

\begin{equation}
    \omega_{QNM} = \Omega_c l - \iota \left( n+\frac{1}{2} \right ) |\lambda|,\label{SSn45}
\end{equation}

where the angular velocity is given by

\begin{equation}
    \Omega_c = \frac{\Dot{\phi}}{\Dot{t}} \Bigr|_{r=r_c} =\sqrt{\frac{f(r_{c})}{r_{c}}}.
\end{equation}

where $r_c$ is the radius of the circular null geodesics. Equation \eqref{SSn45} allows us to write a relation between the QNMs and shadow radius at the eikonal limit \cite{Jusufi:2019ltj}

\begin{equation}
    Re(\omega)=\lim_{l >> 1} \frac{l}{R_{sh}}.\label{ss45}
\end{equation}

Expression \eqref{ss45} is only valid for the large orbital quantum numbers ($l$). But, Konoplya and Stuchik \cite{Konoplya:2017wot} demonstrated that this may not always be the case. The results seen in Table \eqref{st1} are based on Eqs. \eqref{SSn45} and \eqref{ss45}. As can be deduced from Table \eqref{st1}, the results obtained are in agreement with Eq. \eqref{ss45}, however they are inconsistent from the ones obtained for scalar QNMs which are tabulated in Table \eqref{t1}. Nevertheless, one can conclude that the relationship between the shadow radius and QNMs is reliable for $l\gg 1$, which covers the majority of the cases in black hole physics.

\begin{table}
   \centering
   \begin{tabular}{|c|c|c|c|c|c|c|c|}
\hline
$D$ & $r_{ps}$ & $R_{sh}$ & $Re[\omega(r_{ps})]$ & $Re[\omega(R_{sh})]$ & $Im[\omega(r_{ps})]$ & $Im[\omega(R_{sh})]$\\
\hline
4 & 3.00000 & 4.977011373 & 0.1989044067 & 0.2009237924 & 0.09945220335 &
0.08205386720\\
5 & 1.303270425 & 1.837907203 & 0.5386286601 & 0.5440971113 & 0.3808679780 & 0.2867226005\\
6 & 1.060963736 & 1.368415271 & 0.7234276861 & 0.7307723183 & 0.6265067540 & 0.4540265306\\
7 & 0.9933515740 & 1.216002450 & 0.8141015610 & 0.8223667641 & 0.8141015615 & 0.5744345215\\

8 & 0.9764044529 & 1.154930363 & 0.857150798 & 0.8658530696 & 0.9583237370 & 0.6629370490\\

\hline
\end{tabular}
  \caption{The radii of the photon sphere and shadow, and their corresponding QNMs. The physical parameters are chosen as $M=l=1$ and $\Lambda_e=-0.01$.}, \label{st1}
\end{table}

\section{Constraint on Bumblebee Parameter using EHT data obtained for M87* and S\lowercase{gr} A*}\label{sec:cnstraint}
A lot of effort has been paid to evaluating the effects of black holes on the astrophysical environment \cite{Bambi:2017iyh, Bambi:2019xzp, Reynolds:2004qk, AS77}. In particular, we can point to supermassive black holes because they are typically found near the center of galaxies \cite{AS78,AS79}. The supermassive black hole is located at the center of the nearby gigantic elliptical galaxy Messier 87 (M87), also known as M$87^{\ast}$, according to astrophysical measurements cited in \cite{AS80}. The Event Horizon Telescope (EHT) was constructed in order to image the shadow of M$87^{\ast}$ and the supermassive black holes in the heart of the Milky Way (Sgr $A^{\ast}$). Recent investigations have shown that M$87^{\ast}$ has a shadow, as shown by \cite{EventHorizonTelescope:2019dse,AS81,AS82}. The results achieved, as we all know, were fantastic. \\
In this section, we shall constrain the bumblebee parameter by using the data supplied by the EHT for M$87^{\ast}$ and Sgr $A^{\ast}$. As it was reported in \cite{EventHorizonTelescope:2019dse}, angular diameter of the M$87^{\ast}$ black hole shadow is $\theta_\text{M$87^{\ast}$} = 42 \pm 3 \:\mu as$, distance of M$87^{\ast}$ from the Earth is measured as $d_{s}^{M87^{\ast}} = 16.8$ Mpc, and mass of the M$87^{\ast}$ is $M_\text{M87*} = 6.5 \pm 0.90$x$10^9 \: M_\odot$. Similarly, for Sgr $A^{\ast}$ the data for its shadow is given in the latest EHT paper \cite{EventHorizonTelescope:2022wkp}. It is reported that the angular diameter of the Sgr $A^{\ast}$ shadow is $\theta_\text{Sgr $A^{\ast}$} = 48.7 \pm 7 \:\mu as$, distance of the Sgr $A^{\ast}$ from the Earth is $d_{s}^{Sgr A^{\ast}} = 8277\pm33$ pc and mass of the Sgr $A^{\ast}$ black hole is $M_\text{Sgr $A^{\ast}$} = 4.3 \pm 0.013$x$10^6 \: M_\odot$. Now by using those data and following the formula, one can calculate the diameter of the black hole shadow \cite{Bambi:2019tjh},

\begin{equation}
    d_\text{sbh} = \frac{d_{s}  \theta}{M}.
\end{equation}

\begin{figure}[hbt!]
  \includegraphics[width=.49\textwidth]{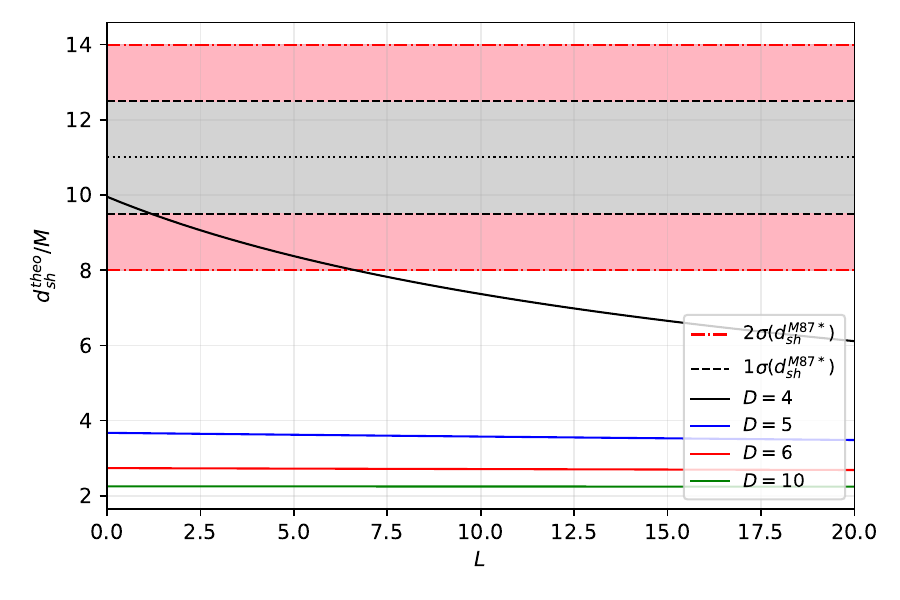}
  \includegraphics[width=.49\textwidth]{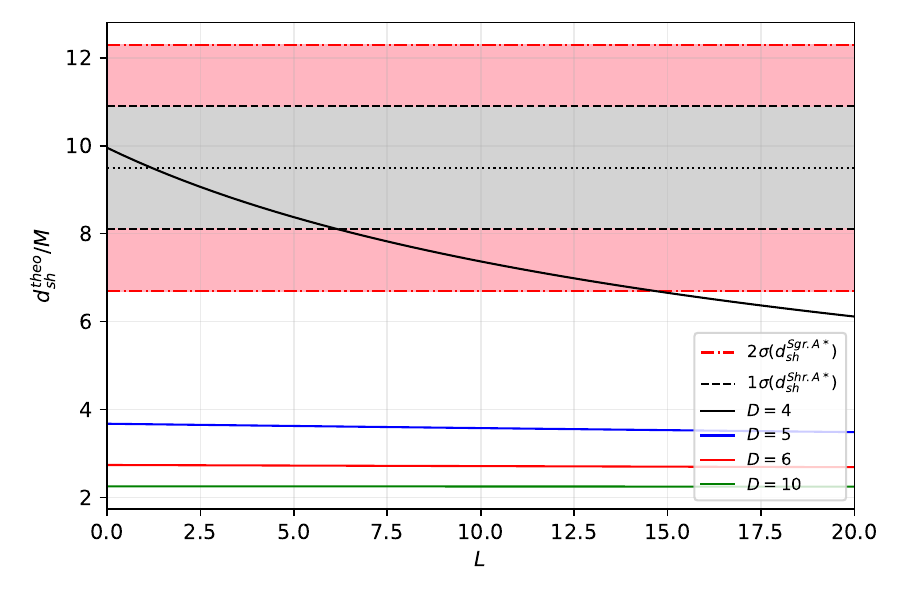}
\caption{The variation of the diameter of the shadow with respect to the bumblebee parameter for different dimensions. Here, we have considered $1\sigma$ and $2\sigma$ uncertainties for the M$87^{\ast}$  (left) and Sgr $A^{\ast}$ (right) for $\Lambda_e=-0.01$.} \label{fig:14a}
\end{figure}

\begin{figure}[hbt!]
  \includegraphics[width=.49\textwidth]{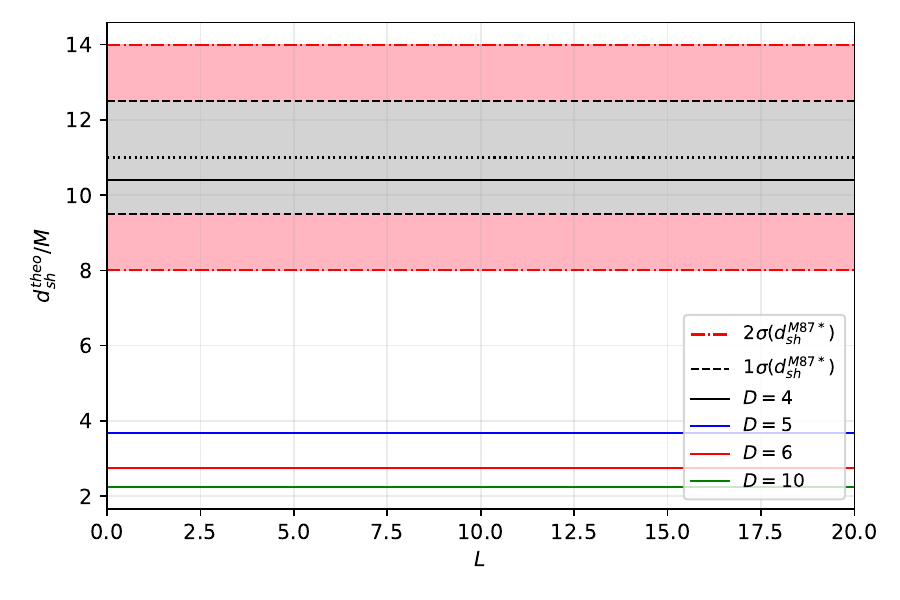}
  \includegraphics[width=.49\textwidth]{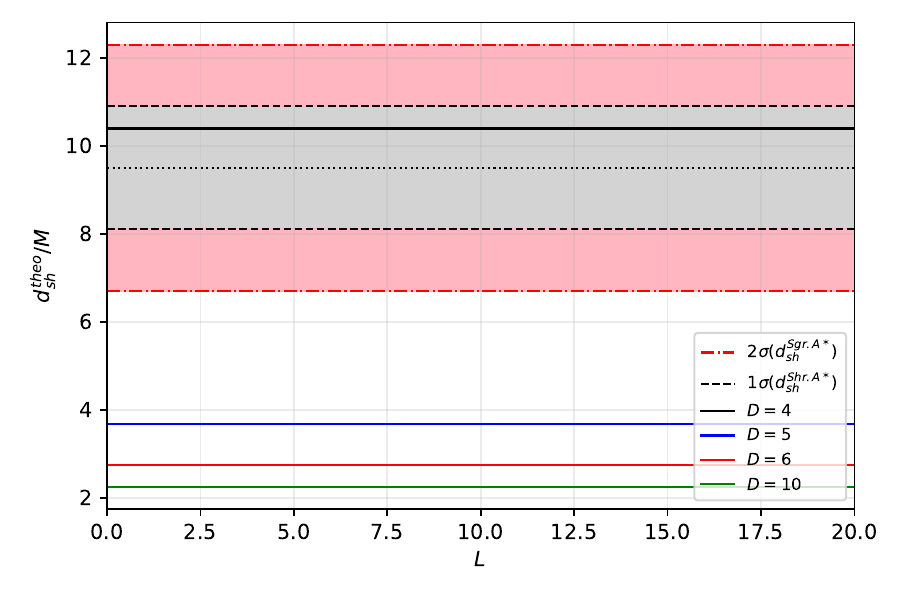}
\caption{The variation of the diameter of the shadow with respect to the bumblebee parameter for different dimensions. Here, we have considered $1\sigma$ and $2\sigma$ uncertainties for the M$87^{\ast}$  (left) and Sgr $A^{\ast}$ (right) for $\Lambda_e=1.11*10^{-52} m^{-2}$.} \label{fig:14b}
\end{figure}

From above, radial diameters of the shadow images for M$87^{\ast}$ and Sgr $A^{\ast}$ can be obtained as $d^\text{M87*}_\text{sbh} = (11 \pm 1.5)M$ and $d^\text{Sgr $A^{\ast}$}_\text{sbh} = (9.5 \pm 1.4)M$, respectively. However, for our metric, the diameter of the black hole shadow can also be calculated from \eqref{92}. Therefore, variation of the diameter of the black hole shadow with the bumblebee parameter for different dimensions can be easily studied as we depicted in Fig. \ref{fig:14a} for $\Lambda_e=-0.01$ and in Fig. \ref{fig:14b} for $\Lambda_e=1.11*10^{-52} m^{-2}$. We have considered $1\sigma$ and $2\sigma$ uncertainties to show the constraints properly. The pertinent figure clearly shows that the bumblebee parameter has a range for $D=4$ that falls within a range of uncertainty, but for higher dimensions, this is not the case for both M$87^{\ast}$ and Sgr $A^{\ast}$ for AdS case. However for the dS case, as we saw in the earlier plots $L$ does not affect the shadow size and therefore for $D=4$, we have all possibility to consider any $L$ value and for higher dimensions, it again falls below the sigma regions. Consequently, in the future, if some observations are made, which suggest a smaller shadow size than what we have observed so far, the current study can thus reveal the importance of the existence of bumblebee gravity.

\section{Conclusions}\label{sec:conclude}

In this work, we have performed a comprehensive discussion on GbFs in higher dimensional AdS/dS black hole spacetimes of the EBGT. The study has provided impressive results in higher dimensions when considering general relativity coupled to the bumblebee gravity. To compute the GbFs, we have considered the scalar and Dirac field perturbations. To analyze the obtained radial wave equations, we have employed the WKB approach up to sixth order to derive the GbFs. The effects of LSB or bumblebee parameter ($L$), cosmological constant ($\Lambda_e$), and dimension ($D$) on the GbFs have been thoroughly investigated, which provide significant impacts on the thermal radiation. It has been observed from Fig. \eqref{FigureS1} that the $4$-dimensional AdS bumblebee black hole has the highest GbF values, whereas the $5$-dimensional black hole has the weakest GbF values. Other higher dimensions ($D>5$) have GbF values between the fourth and fifth dimensions, though. The GbF values drop as the dimension rises in $D>5$ dimensional black holes. On the other hand, the bosonic GbFs of dS higher dimensional bumblebee black holes are shown in Fig. \eqref{FigureS2}, which decrease regularly with increasing dimension. It is also understood that at higher dimensions, the LSB effect on the bosonic GbFs of the AdS higher dimensional bumblebee black holes is very weak compared to the dimension effect. The GbFs of the dS higher dimensional bumblebee black holes are effectively reduced by the increasing LSB, though. 

Fig. \eqref{fig:S8a} shows the fermionic GbF behaviors of the higher ($D\geq4$) dimensional AdS/dS bumblebee black holes against the change in dimension and the LSB parameter. In general, irrespective of being a bosonic or fermionic perturbation, the most important findings obtained from Figs. \ref{FigureS1}, \ref{FigureS2}, and \ref{fig:S8a} are that the GbFs of the $4$-dimensional dS/AdS bumblebee black hole in the EBGT theory are higher than those for $D>4$ black holes. Namely, GbFs drastically reduce with the increasing dimensions, which means that the probability for detecting the thermal radiation of the higher dimensional ds/AdS black holes in the EBGT gets lower with $D > 4$. We have also noticed that the GbFs reach to $1$ quicker in the scalar field perturbations compared to the fermionic field perturbations. This indicates that the bosonic thermal radiations can more likely reach to spatial infinity in comparison to the fermionic thermal emission.

Bosonic and fermionic QNMs with  $l=0$ and $l=1$ cases of the higher dimensional dS/AdS bumblebee black holes are tabulated in Tables \eqref{t1}  an \eqref{t2}, respectively. We have inferred from those tables that both real and imaginary parts of the bosonic and fermionic QNMs increase with growing dimensionality. Almost the same behaviors obtained for the rising $L$ parameter except for the damping mode are obtained at $D=6$ and $D=8$ dimensions for bosons and fermions, respectively. We have also extended our investigation to find a direct link between the QNMs frequencies and the shadows of the bumblebee black holes. The shadows of the higher dimensional ds/AdS bumblebee black holes have been studied in terms of null geodesics and spherical photon orbits. By changing the dimensionality, the black holes' shadow radii have been depicted and analyzed. It is clearly shown in Fig. \ref{sh1} (left) that an increasing dimension ($D\geq4$) decreases the radius of the black hole's shadow.  We have also shown that the latter result can also be obtained by means of the real part of the QNMs frequencies, which are valid in the eikonal limit. Finally, we have considered the null geodesics and obtained the black hole shadow radius with different dimension. Then, we have exhibited the outcomes in Fig. \eqref{sh2} (left). Similarly, we have shown the variation of the shadow radius with the bumblebee parameter ($L$) in Fig. \ref{sh1} (right) for dimension $D=4$. It is clear that shadow radius decreases with increasing the $L$. The same behaviour is depicted in the celestial coordinate in Fig. \ref{sh2} (right). In the sequel, we have examined the variation of the black hole shadow diameter with the bumblebee parameter based on the data of real black holes ( M$87^{\ast}$ and Sgr $A^{\ast}$ (see Fig. \eqref{fig:14a}). At the end of the day, we have shown that the bumblebee parameter has a reducing effect on the diameter of the black hole shadow. This result can be used as a tool to indirectly prove the existence of the bumblebee gravity theory.

In recent times, the search for proof of the existence of bumblebee gravity has has gained momentum. Especially, the research of Gu et al. \cite{Gu:2022grg} based on real black hole X-ray data is quite remarkable. In this context, this study will contribute to the examination of EBGT with possible optical and wave observations to be made in the future. Our research can be expanded to charged bumblebee black holes, which will require taking into account the bumblebee electrodynamics \cite{Seifert:2009gi,Hernaski:2014jsa,Escobar:2017fdi}. A possible charged bumblebee black hole solution will allow us to examine the EBGT with linear/non-linear electrodynamics theories \cite{Paula:2020yfr} and we will likely obtain more detailed results on the thermal radiation (GbFs), gravitational ringing (QNMs), and optical observations (lensing and shadow). This
is the next stage of study that interests us. 

\section*{Acknowledgements}
We would like to thank Marco Schreck and Ali \"{O}vg\"{u}n for useful conversations and discussions. We are also thankful to the Editor and Referee for their constructive suggestions and comments.

\bibliographystyle{apsrev}

\begin{thebibliography}{20}
\bibitem{ParticleDataGroup:2012pjm}
J.~Beringer \textit{et al.} [Particle Data Group],
Phys. Rev. D \textbf{86}, 010001 (2012).

\bibitem{Weinberg:1988cp}
S.~Weinberg,
Rev. Mod. Phys. \textbf{61}, 1 (1989).

\bibitem{Hehl:1994ue}
F.~W.~Hehl, J.~D.~McCrea, E.~W.~Mielke, and Y.~Ne'eman,
Phys. Rept. \textbf{258}, 1 (1995).

\bibitem{Bailey:2006fd}
Q.~G.~Bailey and V.~A.~Kostelecky,
Phys. Rev. D \textbf{74}, 045001 (2006).

\bibitem{Bluhm:2004ep}
R.~Bluhm and V.~A.~Kostelecky,
Phys. Rev. D \textbf{71}, 065008 (2005).

\bibitem{Kostelecky:2010ze}
A.~V.~Kostelecky and J.~D.~Tasson,
Phys. Rev. D \textbf{83}, 016013 (2011).

\bibitem{ParticleDataGroup:2018ovx}
M.~Tanabashi \textit{et al.} [Particle Data Group],
Phys. Rev. D \textbf{98}, 030001 (2018).

\bibitem{Copeland:2006wr}
E.~J.~Copeland, M.~Sami, and S.~Tsujikawa,
Int. J. Mod. Phys. D \textbf{15}, 1753 (2006).

\bibitem{DeWitt:1967yk}
B.~S.~DeWitt,
Phys. Rev. \textbf{160}, 1113 (1967).

\bibitem{Aharony:1999ti}
O.~Aharony, S.~S.~Gubser, J.~M.~Maldacena, H.~Ooguri, and Y.~Oz,
Phys. Rept. \textbf{323}, 183 (2000).

\bibitem{Jungman:1995df}
G.~Jungman, M.~Kamionkowski, and K.~Griest,
Phys. Rept. \textbf{267}, 195 (1996).

\bibitem{Reyes:2021cpx}
C.~M.~Reyes and M.~Schreck,
Phys. Rev. D \textbf{104}, no.12, 124042 (2021)
[arXiv:2105.05954 [gr-qc]].

\bibitem{Liberati:2009pf}
S.~Liberati and L.~Maccione,
Ann. Rev. Nucl. Part. Sci. \textbf{59}, 245 (2009).

\bibitem{Kostelecky:1988zi}
V.~A.~Kostelecky and S.~Samuel,
Phys. Rev. D \textbf{39}, 683 (1989).

\bibitem{Jacobson:2000xp}
T.~Jacobson and D.~Mattingly,
Phys. Rev. D \textbf{64}, 024028 (2001).

\bibitem{Jacobson:2004ts}
T.~Jacobson and D.~Mattingly,
Phys. Rev. D \textbf{70}, 024003 (2004).

\bibitem{Carroll:2001ws}
S.~M.~Carroll, J.~A.~Harvey, V.~A.~Kostelecky, C.~D.~Lane, and T.~Okamoto,
Phys. Rev. Lett. \textbf{87}, 141601 (2001).

\bibitem{Mocioiu:2000ip}
I.~Mocioiu, M.~Pospelov, and R.~Roiban,
Phys. Lett. B \textbf{489}, 390 (2000).

\bibitem{Ferrari:2006gs}
A.~F.~Ferrari, M.~Gomes, J.~R.~Nascimento, E.~Passos, A.~Y.~Petrov, and A.~J.~da Silva,
Phys. Lett. B \textbf{652}, 174 (2007).

\bibitem{Gambini:1998it}
R.~Gambini and J.~Pullin,
Phys. Rev. D \textbf{59}, 124021 (1999).

\bibitem{Ellis:1999uh}
J.~R.~Ellis, N.~E.~Mavromatos, and D.~V.~Nanopoulos,
Gen. Rel. Grav. \textbf{32}, 127 (2000).

\bibitem{Burgess:2002tb}
C.~P.~Burgess, J.~M.~Cline, E.~Filotas, J.~Matias, and G.~D.~Moore,
JHEP \textbf{03}, 043 (2002).

\bibitem{Frey:2003jq}
A.~R.~Frey,
JHEP \textbf{04}, 012 (2003).

\bibitem{Fernando:2014gda}
S.~Fernando and T.~Clark,
Gen. Rel. Grav. \textbf{46}, 1834 (2014).

\bibitem{Reyes:2022mvm}
C.~M.~Reyes and M.~Schreck,
[arXiv:2202.11881 [hep-th]].


\bibitem{Bluhm:2007bd}
R.~Bluhm, S.~H.~Fung, and V.~A.~Kostelecky,
Phys. Rev. D \textbf{77}, 065020 (2008).

\bibitem{Bertolami:2005bh}
O.~Bertolami and J.~Paramos,
Phys. Rev. D \textbf{72} 044001 (2005).

\bibitem{Casana:2017jkc}
R.~Casana, A.~Cavalcante, F.~P.~Poulis, and E.~B.~Santos,
Phys. Rev. D \textbf{97}, 104001 (2018).

\bibitem{sarak1}
S.~Kanzi and \.I.~Sakall\i{},
Nucl. Phys. B \textbf{946}, 114703 (2019).

\bibitem{sarak2}
S.~Kanzi and \.I.~Sakall\i{},
Eur. Phys. J. C \textbf{81}, 501 (2021).

\bibitem{sarak3}
S.~Kanzi, S.~H.~Mazharimousavi, and \.I.~Sakall\i{},
Annals Phys. \textbf{422}, 168301 (2020).

\bibitem{sarak4}
A.~Al-Badawi, \.I.~Sakall\i{}, and S.~Kanzi,
Annals Phys. \textbf{412}, 168026 (2020).

\bibitem{sarak5}
S.~Kanzi and \.I.~Sakall\i{},
Eur. Phys. J. Plus \textbf{137}, 14 (2022).

\bibitem{sarak6}
\.I.~Sakall\i{} and S.~Kanzi,
Annals Phys. \textbf{439}, 168803 (2022).

\bibitem{sarak7}
A.~Al-Badawi, S.~Kanzi, and \.I.~Sakall\i{},
Eur. Phys. J. Plus \textbf{137}, 94 (2022).

\bibitem{sarak8}
A.~Al-Badawi, S.~Kanzi, and \.I.~Sakall\i{},
[arXiv:2203.04140 [hep-th]].


\bibitem{Ding:2020kfr}
C.~Ding and X.~Chen,
Chin. Phys. C \textbf{45} 025106 (2021).

\bibitem{Maluf:2022knd}
R.~V.~Maluf and C.~R.~Muniz,
Eur. Phys. J. C \textbf{82}, 94 (2022).

\bibitem{Kanzi:2022vhp}
S.~Kanzi and \.I.~Sakall\i{},
Eur. Phys. J. C \textbf{82}, 93 (2022).

\bibitem{Ding:2019mal}
C.~Ding, C.~Liu, R.~Casana, and A.~Cavalcante,
Eur. Phys. J. C \textbf{80}, 178 (2020).

\bibitem{Jha:2020pvk}
S.~K.~Jha and A.~Rahaman,
Eur. Phys. J. C \textbf{81}, 345 (2021).

\bibitem{Poulis:2021nqh}
F.~P.~Poulis and M.~A.~C.~Soares,
[arXiv:2112.04040 [gr-qc]].

\bibitem{Ding:2022qcy}
C.~Ding, Y.~Shi, J.~Chen, Y.~Zhou, and C.~Liu,
[arXiv:2201.06683 [gr-qc]].

\bibitem{Chandrasekhar:1984siy}
S.~Chandrasekhar,
Fundam. Theor. Phys. \textbf{9}, 5 (1984).

\bibitem{Ovgun:2017dvs}
A.~\"Ovg\"un, \.I.~Sakall\i{}, and J.~Saavedra,
Chin. Phys. C \textbf{42}, 105102 (2018).

\bibitem{Jusufi:2017trn}
K.~Jusufi, I.~Sakall\i{}, and A.~Ovg\"un,
Gen. Rel. Grav. \textbf{50}, 10 (2018).

\bibitem{Sakalli:2021dxd}
\.I.~Sakalli and G.~T.~Hyusein,
Turk. J. Phys. \textbf{45}, 43 (2021).

\bibitem{Sakalli:2016fif}
I.~Sakalli,
Phys. Rev. D \textbf{94}, 084040 (2016).

\bibitem{Sakalli:2018nug}
\.I.~Sakall\i{}, K.~Jusufi, and A.~\"Ovg\"un,
Gen. Rel. Grav. \textbf{50}, 125 (2018).

\bibitem{Schutz:1985km}
B.~F.~Schutz and C.~M.~Will,
Astrophys. J. Lett. \textbf{291}, L33-L36 (1985).

\bibitem{Iyer:1986np}
S.~Iyer and C.~M.~Will,
Phys. Rev. D \textbf{35}, 3621 (1987).

\bibitem{Iyer:1986nq}
S.~Iyer,
Phys. Rev. D \textbf{35}, 3632 (1987).

\bibitem{Konoplya:2011qq}
R.~A.~Konoplya and A.~Zhidenko,
Rev. Mod. Phys. \textbf{83}, 793 (2011).
\bibitem{Leaver:1985ax}
E.~W.~Leaver,
Proc. Roy. Soc. Lond. A \textbf{402}, 285 (1985).

\bibitem{izmashoon}H.-J. Blome and B. Mashhoon, 
 Phys. Lett. A \textbf{100}, 231 (1984).
 
 \bibitem{Ovgun:2019yor}
A.~\"Ovg\"un, \.I.~Sakall\i{}, and H.~Mutuk,
Int. J. Geom. Meth. Mod. Phys. \textbf{18}, 2150154 (2021).
 
 \bibitem{Kokkotas:1999bd}
K.~D.~Kokkotas and B.~G.~Schmidt,
Living Rev. Rel. \textbf{2}, 2 (1999).

\bibitem{Berti:2009kk}
E.~Berti, V.~Cardoso, and A.~O.~Starinets,
Class. Quant. Grav. \textbf{26}, 163001 (2009).

\bibitem{Sakalli:2022xrb}
\.I.~Sakalli and S.~Kanzi,
Turk. J. Phys. \textbf{46}, 51 (2022).

\bibitem{Konoplya:2003ii}
R.~A.~Konoplya,
Phys. Rev. D \textbf{68}, 024018 (2003).

\bibitem{EHT:2019nmr}
C.~Goddi \textit{et al.} [EHT],
The Messenger \textbf{177}, 25 (2019).

\bibitem{EventHorizonTelescope:2019dse}
K.~Akiyama \textit{et al.} [Event Horizon Telescope],
Astrophys. J. Lett. \textbf{875}, L1 (2019).

\bibitem{EventHorizonTelescope:2019ths}
K.~Akiyama \textit{et al.} [Event Horizon Telescope],
Astrophys. J. Lett. \textbf{875}, L4 (2019).

\bibitem{EventHorizonTelescope:2019pgp}
K.~Akiyama \textit{et al.} [Event Horizon Telescope],
Astrophys. J. Lett. \textbf{875}, L5 (2019).

\bibitem{EventHorizonTelescope:2022xnr}
K.~Akiyama \textit{et al.} [Event Horizon Telescope],
Astrophys. J. Lett. \textbf{930}, no.2, L12 (2022).

\bibitem{Dokuchaev:2020wqk}
V.~I.~Dokuchaev and N.~O.~Nazarova,
Universe \textbf{6}, 154 (2020).

\bibitem{Synge:1966okc}
J.~L.~Synge,
Mon. Not. Roy. Astron. Soc. \textbf{131}, 463 (1966).

\bibitem{Bardeen:1973tla}
J. M. Bardeen, 
Les Houches Summer School of Theoretical Physics: Black Holes, 215-240 (1973).

\bibitem{Neves:2019lio}
J.~C.~S.~Neves,
Eur. Phys. J. C \textbf{80}, 343 (2020).

\bibitem{Vagnozzi:2019apd}
S.~Vagnozzi and L.~Visinelli,
Phys. Rev. D \textbf{100}, no.2, 024020 (2019).
\bibitem{Kumar:2020yem}
R.~Kumar, A.~Kumar, and S.~G.~Ghosh,
Astrophys. J. \textbf{896}, 89 (2020).


\bibitem{Uniyal:2022vdu}
A.~Uniyal, R.~C.~Pantig, and A.~\"Ovg\"un,
[arXiv:2205.11072 [gr-qc]].

\bibitem{Kuang:2022xjp}
X.~M.~Kuang and A.~\"Ovg\"un,
[arXiv:2205.11003 [gr-qc]].


\bibitem{Boulware:1985wk}
D.~G.~Boulware and S.~Deser,
Phys. Rev. Lett. \textbf{55}, 2656 (1985).

\bibitem{Harmark:2007jy}
T.~Harmark, J.~Natario, and R.~Schiappa,
Adv. Theor. Math. Phys. \textbf{14}, 727 (2010).

\bibitem{Sakalli:2011zz}
I.~Sakalli,
Int. J. Mod. Phys. A \textbf{26}, 2263 (2011);
[erratum: Int. J. Mod. Phys. A \textbf{28}, 1392002 (2013)].

\bibitem{Cho:2007zi}
H.~T.~Cho, A.~S.~Cornell, J.~Doukas, and W.~Naylor,
Phys. Rev. D \textbf{75}, 104005 (2007). 

\bibitem{Chakrabarti:2008xz}
S.~K.~Chakrabarti,
Eur. Phys. J. C \textbf{61}, 477 (2009).


\bibitem{Das:1996we}
S.~R.~Das, G.~W.~Gibbons, and S.~D.~Mathur,
Phys. Rev. Lett. \textbf{78}, 417 (1997).


\bibitem{Gibbons:1993hg}
G.~W.~Gibbons and A.~R.~Steif,
Phys. Lett. B \textbf{314}, 13 (1993).

\bibitem{Camporesi:1995fb}
R.~Camporesi and A.~Higuchi,
J. Geom. Phys. \textbf{20}, 1 (1996).


\bibitem{Gullu:2020qzu}
\.I.~G\"ull\"u and A.~\"Ovg\"un,
Annals Phys. \textbf{436}, 168721 (2022). 

\bibitem{Luminet:1979nyg}
J.~P.~Luminet,
Astron. Astrophys. \textbf{75}, 228 (1979).

\bibitem{Perlick:2021aok}
V.~Perlick and O.~Y.~Tsupko,
Phys. Rept. \textbf{947}, 1 (2022).

\bibitem{Konoplya:2019sns}
R.~A.~Konoplya,
Phys. Lett. B \textbf{795}, 1 (2019).

\bibitem{Sakalli:2014bfa}
I.~Sakalli, A.~Ovgun and S.~F.~Mirekhtiary,
Int. J. Geom. Meth. Mod. Phys. \textbf{11}, 1450074 (2014).

\bibitem{Mirekhtiary:2019ugi}
S.~F.~Mirekhtiary and I.~Sakalli,
Teor. Mat. Fiz. \textbf{198} (2019) no.3, 523-531


\bibitem{Du:2022hom}
Y.~Z.~Du, H.~F.~Li, X.~N.~Zhou, W.~Q.~Guo and R.~Zhao,
[arXiv:2206.14382 [hep-th]].

\bibitem{Stefanov:2010xz}
I.~Z.~Stefanov, S.~S.~Yazadjiev, and G.~G.~Gyulchev,
Phys. Rev. Lett. \textbf{104}, 251103 (2010). 

\bibitem{Jusufi:2019ltj}
K.~Jusufi,
Phys. Rev. D \textbf{101}, 084055 (2020).

\bibitem{Cuadros-Melgar:2020kqn}
B.~Cuadros-Melgar, R.~D.~B.~Fontana, and J.~de Oliveira,
Phys. Lett. B \textbf{811}, 135966 (2020). 

\bibitem{Moura:2021eln}
F.~Moura and J.~Rodrigues,
Phys. Lett. B \textbf{819}, 136407 (2021). 

\bibitem{Liu:2020ola}
C.~Liu, T.~Zhu, Q.~Wu, K.~Jusufi, M.~Jamil, M.~Azreg-A\"\i{}nou, and A.~Wang,
Phys. Rev. D \textbf{101}, 084001 (2020); 
[erratum: Phys. Rev. D \textbf{103}, 089902 (2021).

\bibitem{Cardoso:2008bp}
V.~Cardoso, A.~S.~Miranda, E.~Berti, H.~Witek, and V.~T.~Zanchin,
Phys. Rev. D \textbf{79}, 064016 (2009). 

\bibitem{Konoplya:2017wot}
R.~A.~Konoplya and Z.~Stuchl\'\i{}k,
Phys. Lett. B \textbf{771}, 597 (2017).

\bibitem{Gu:2022grg}
J.~Gu, S.~Riaz, A.~B.~Abdikamalov, D.~Ayzenberg, and C.~Bambi,
[arXiv:2206.14733 [gr-qc]].

\bibitem{Seifert:2009gi}
M.~D.~Seifert,
Phys. Rev. D \textbf{81}, 065010 (2010).

\bibitem{Hernaski:2014jsa}
C.~Hernaski,
Phys. Rev. D \textbf{90}, no.12, 124036 (2014).

\bibitem{Escobar:2017fdi}
C.~A.~Escobar and A.~Mart\'\i{}n-Ruiz,
Phys. Rev. D \textbf{95}, 095006 (2017).

\bibitem{Paula:2020yfr}
M.~A.~A.~Paula, L.~C.~S.~Leite, and L.~C.~B.~Crispino,
Phys. Rev. D \textbf{102}, 104033 (2020).

\bibitem{Bambi:2017iyh}
C.~Bambi,
Annalen Phys. \textbf{530}, 1700430 (2018)

\bibitem{Bambi:2019xzp}
C.~Bambi,
PoS \textbf{MULTIF2019}, 028 (2020)
[arXiv:1906.03871 [astro-ph.HE]].

\bibitem{Reynolds:2004qk}
C.~S.~Reynolds, L.~W.~Brenneman and D.~Garofalo,
Astrophys. Space Sci. \textbf{300}, 71-79 (2005)
[arXiv:astro-ph/0410116 [astro-ph]].

\bibitem{AS77}
A.~E.~Reines,
Nature \textbf{513}, 322-323 (2014).

\bibitem{AS78}
D.~Lynden-Bell,
Nature \textbf{223}, 690-694 (1969).

\bibitem{AS79}
J.~Kormendy and K.~Gebhardt,
AIP Conf. Proc. \textbf{586}, no.1, 363-381 (2001)
[arXiv:astro-ph/0105230 [astro-ph]].

\bibitem{AS80}
K. Gebhardt, J. Adams, D. Richstone, T. R. Lauer,
S. M. Faber, K. Gultekin et al.,
Astrophys. J. \textbf{729}, 119 (1954).

\bibitem{EventHorizonTelescope:2019dse}
K.~Akiyama \textit{et al.} [Event Horizon Telescope],
Astrophys. J. Lett. \textbf{875} (2019), L1.
[arXiv:1906.11238 [astro-ph.GA]].

\bibitem{AS81}
K.~Akiyama \textit{et al.} [Event Horizon Telescope],
Astrophys. J. Lett. \textbf{875} (2019), L2.

\bibitem{AS82}
K.~Akiyama \textit{et al.} [Event Horizon Telescope],
Astrophys. J. Lett. \textbf{875} (2019), L3.


\bibitem{EventHorizonTelescope:2022wkp}
K.~Akiyama \textit{et al.} [Event Horizon Telescope],
Astrophys. J. Lett. \textbf{930} (2022) no.2, L12
doi:10.3847/2041-8213/ac6674

\bibitem{Bambi:2019tjh}
C.~Bambi, K.~Freese, S.~Vagnozzi and L.~Visinelli,
Phys. Rev. D \textbf{100} (2019) no.4, 044057
doi:10.1103/PhysRevD.100.044057
[arXiv:1904.12983 [gr-qc]].

\end{thebibliography}

\end{document}